\documentclass[manuscript]{aastex}

\slugcomment{Not to appear in Nonlearned J., 45.}

\shorttitle{MRI with CR-diffusion}
\shortauthors{Kuwabara and Ko}

\begin{document}


\title{ANALYSIS OF THE MAGNETO ROTATIONAL INSTABILITY WITH \\
       THE EFFECT OF COSMIC-RAY DIFFUSION}



\author{Takuhito Kuwabara}
\affil{Computational Science and Engineering Division I,
       AdvanceSoft Corporation, 4-3, Kanda Surugadai, Chiyoda-ku,
       Tokyo 101-0062, Japan}
\email{kuwabrtk@gmail.com}


\and

\author{Chung-Ming Ko}
\affil{Department of Physics, Institute of Astronomy and Center
       for Complex Systems, National Central University,
       Jhongli, Taiwan 320, Republic of China}
\email{cmko@astro.ncu.edu.tw}






\begin{abstract}

We present the results obtained from linear stability analysis and
2.5-dimensional magnetohydrodynamic (MHD) simulations of the
magnetorotational instability (MRI), including the effects of cosmic
rays (CRs). We took into account of the CR diffusion along the
magnetic field but neglect the cross-field-line diffusion. Two
models are considered in this paper: shearing box model and
differentially rotating cylinder model. We studied how MRI is
affected by the initial CR pressure (i.e., energy) distribution. In
the shearing box model, the initial state is uniform distribution.
Linear analysis shows that the growth rate of MRI does not depend on
the value of CR diffusion coefficient.
In the differentially rotating cylinder model, the initial state is a constant angular momentum polytropic disk
threaded by weak uniform vertical magnetic field. Linear analysis shows that the growth rate of MRI becomes larger
if the CR diffusion coefficient is larger.
Both results are confirmed by MHD simulations.
The MHD simulation results show that the outward movement of matter by the growth of MRI is not impeded by the
CR pressure gradient, and the centrifugal force which acts to the concentrated matter becomes larger.
Consequently, the growth rate of MRI is increased.
On the other hand, if the initial CR pressure is uniform, then the growth rate of the MRI barely depends on
the value of the CR diffusion coefficient.
\end{abstract}






\section{INTRODUCTION}

Magnetic field, an important component of the interstellar medium (ISM),
is thought to be a key player in various active astrophysical phenomena.
However, the dynamical role of cosmic ray (CR) (another component of ISM) in astrophysical
activities has been underrated for quite a long time, although the energy density of CRs
is of the same order as that of magnetic field and turbulent gas motions
\citep[e.g.,][]{parker1969,ginzburg1976,ferriere2001}.
Still some effort have been made over the years.
The most convenient way to study the effect of CRs on plasma flow is to describe the system
as a multi-fluid system where plasma and CRs are considered as fluids
\citep[e.g.,][]{Drury1981, Axford1982, Webb1986, Webb1987, Ko1997}.
One may also consider the self-excited waves as fluids in the CR-plasma system
\citep[e.g.,][]{McKenzie1982, Ko1991a, Ko1991b, Ko1991c, Ko1992, Ko2001}.
The system exhibits some unique instabilities, e.g., squeezing instability
\citep[see e.g.,][]{Drury1986, Zank1987, Zank1990, Kang1992, Zank1993},
and magneto-acoustic instability
\citep[see e.g.,][]{McKenzie1984, Zank1989, Ko1994, Lo2007, Ko2009}.

The influence of CRs on various instabilities has been studied by means of linear analysis and
MHD simulations, for instance, the evolution of Parker instability
\citep[][]{parker1966, hanasz1997, hanasz2000, ryu, kuwabara2004, Lo2011}, Parker-Jeans instability
\citep[][]{Kuznetsov1983, kuwabara2006},
magneto-rotational instability \citep[][]{khajenabi2012}, Kelvin-Helmholtz instability
\citep[][]{suzuki2014}, and also galactic dynamo \citep[][]{parker1992, hanasz2004, hanasz2009}.
The results of these works showed that in some cases the growth rate has some intriguing dependence on
the cosmic ray pressure and the coupling of CR and thermal plasma (i.e., the cosmic ray diffusion
coefficient).
For example, while the cosmic ray pressure may effectively enhance the Parker and Kelvin-Helmholtz instabilities,
small diffusion coefficient can impede the growth \citep[][]{suzuki2014}.
Moreover, the diffusion coefficient may determine the fragmentation direction of Parker-Jeans instability
\citep[][]{kuwabara2006}.

Magneto-rotational instability (MRI) is an important mechanism in differentially rotating astrophysical objects
with magnetic fields.
\citet{balbus1991} and \citet{hawley1991} showed that local and extremely powerful instability in a differentially
rotating systems with a weak magnetic field destabilize the systems strongly.
As MRI occurs in accretion disk, the magnetic energy is amplified inside the disk, and angular momentum transfer
takes place, which is important for obtaining high enough accretion rate to explain observations.
The efficiency of angular momentum transport can be estimated from the saturation level of the magnetic energy,
and \citet{sano1998} showed that the saturation level of MRI using the resistive MHD simulations.

\citet{khajenabi2012} studied the influence of CRs on MRI in the case of dominant toroidal magnetic field in the linear regime,
and showed that the CR pressure enhanced the growth of MRI and the diffusion of CRs suppressed the growth of MRI.
In this work, we analyze the case of dominant poloidal magnetic field by linear perturbation analysis and MHD numerical simulations.
We arrive at a somewhat different conclusion.
We find similar enhancement of MRI by CR pressure as in \citet{khajenabi2012}.
However, we notice that diffusion of CRs enhances the growth of MRI as well.
This may be alluded to the fact that we are using non-uniform initial equilibrium state
and different orientation of the magnetic field.
Similar result is observed in Parker or Parker-Jeans instabilities with CRs
\citep[][]{kuwabara2004,kuwabara2006}.

This paper is organized as follows.
In section \ref{sec:models} we describe the two-fluid model of CR-plasma system.
In this section we present the governing equations of the shearing box model and the rotating cylinder model,
and their equilibrium models.
In section \ref{sec:linear}, the linear stability analysis and its results of the two models are presented, and
in section \ref{sec:simulation} the results of MHD simulations are presented.
section \ref{sec:summary} provides a summary and discussion.



\section{MODELS}\label{sec:models}

We study the MRI in differentially rotating disk in the context of the two-fluid CR-plasma system.
CR is considered as a massless fluid but with significant pressure.
The CR fluid is couple to the other fluid, thermal plasma,
through the embedded magnetic irregularities or hydromagnetic waves.
To a first approximation, the effect of waves is contained in the hydrodynamical diffusion coefficient
of CR and this diffusion coefficient serves as the coupling between the two fluids.
The system is governed by the total mass, momentum and energy equations for the thermal plasma,
cosmic ray and magnetic field.

The cosmic ray energy equation describes the energy transfer between the plasma and CR.
In this work, we ignore the cross-field-line diffusion of CRs, as in many cases
the ratio of the perpendicular diffusion coefficient to the parallel one is quite small,
$0.02\sim 0.04$ \citep[e.g.,][]{giacalone, ryu}.
Moreover, ideal MHD is assumed in this work.
The cases for cross-field-line diffusion and non-ideal MHD will be considered in subsequent work.

The set of governing equations in rotating frame is:
\begin{equation}\label{mass-total}
  {\partial\rho\over\partial t}+{\bf\nabla}\cdot(\rho{\bf V})=0\,,
\end{equation}
\begin{equation}\label{momentum-total}
  {\!\partial\over\partial t}\left(\rho{\bf V}\right)+
  {\bf\nabla}\cdot\left[\rho{\bf V}{\bf V}+\left(P_{\rm g}+P_{\rm c}+{B^2\over 2\mu_0}\right){\bf I}-
  {{\bf B}{\bf B}\over\mu_0}\right]
  +\rho\left[2{\bf\Omega}\times{\bf V}+{\bf\Omega}\times\left({\bf\Omega}\times{\bf r}\right)-{\bf g}\right]=0\,,
\end{equation}
\begin{equation}\label{plasma-energy}
  {\partial P_{\rm g}\over\partial t}+{\bf V}\cdot{\bf\nabla}P_{\rm g}
  +\gamma_{\rm g}P_{\rm g}{\bf\nabla}\cdot{\bf V}=0\,,
\end{equation}
\begin{equation}\label{cr-energy}
  {\partial P_{\rm c}\over\partial t}+{\bf V}\cdot{\bf\nabla}P_{\rm c}
  +\gamma_{\rm c}P_{\rm c}{\bf\nabla}\cdot{\bf V}
  -{\bf\nabla}\cdot\left(\kappa_{\|}{{\bf B}\over B}{{\bf B}\over B}\cdot{\bf\nabla}P_{\rm c}\right)=0\,,
\end{equation}
\begin{equation}\label{mag-energy}
  {\partial{\bf B}\over\partial t}-{\bf\nabla}\cdot({\bf V}\times{\bf B})=0\,,
\end{equation}
where $\rho$ and ${\bf V}$ are the plasma density and velocity,
$P_{\rm g}$ and $P_{\rm c}$ are the thermal pressure and the CR pressure,
$\gamma_{\rm g}$ and $\gamma_{\rm c}$ are the polytropic index for the plasma and the CRs
(i.e., the energy densities of the thermal plasma and CR are given by
$E_{\rm th}=P_{\rm g}/(\gamma_{\rm g}-1)$ and $E_{\rm c}=P_{\rm c}/(\gamma_{\rm c}-1)$),
${\bf B}$ is the magnetic field, $B$ is the magnitude of magnetic field strength,
$\kappa_{\|}$ is the CR diffusion coefficient along the magnetic field,
$I$ is the unit tensor,
and ${\bf g}$ is gravity and ${\bf\Omega}$ is the angular velocity of the rotating frame.
Equation~(\ref{mag-energy}) is the Faraday's induction equation. The inner product of this equation with
${\bf B}$ gives the energy equation for the magnetic field.

In the following we adopt two models for the differentially rotating disk:
the shearing box model and the differentially rotating cylinder model.

\subsection{Shearing box}\label{sec:shear1}
\begin{figure}
\epsscale{.80}
\plotone{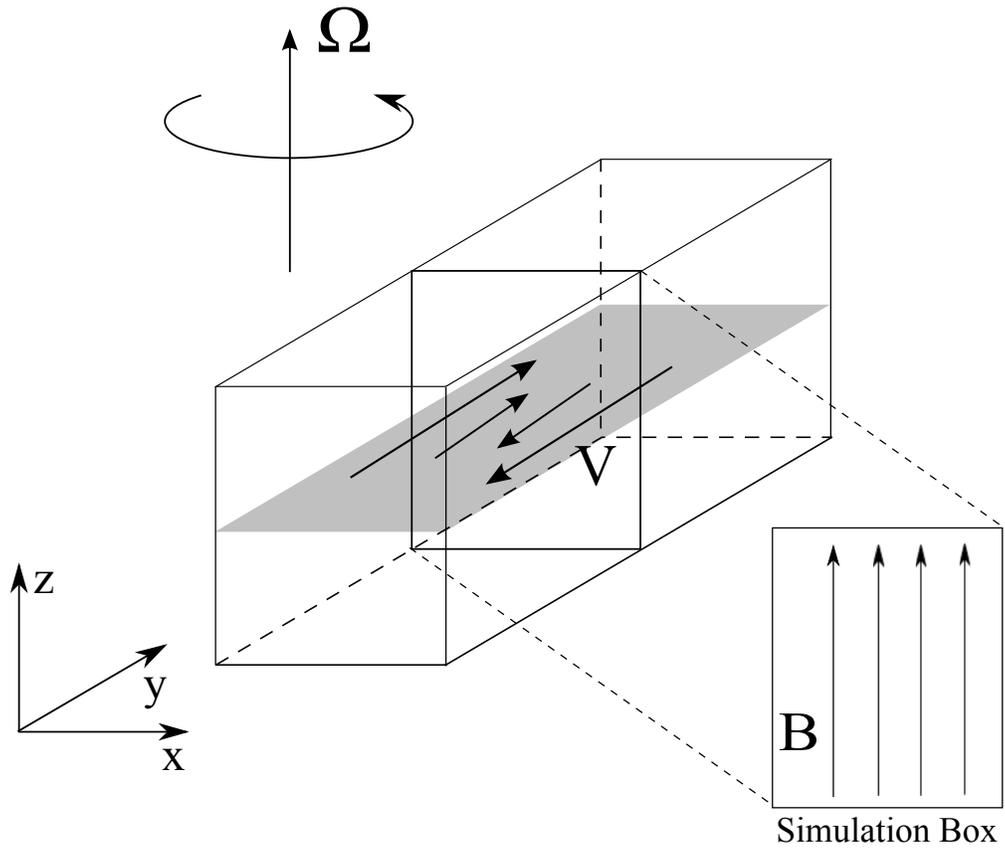}
\caption{Schematic picture of the shearing box model.}
\label{fig1}
\end{figure}

We consider a two-dimension shearing box in a rotating frame. We choose the local Cartesian coordinates $(x,y,z)$,
where $\hat{\bf e}_x$ is the radial direction, and the angular velocity of the rotating frame is
$\bf\Omega=\Omega\hat{\bf e}_z$ (see Figure~\ref{fig1}).
The centrifugal force term together with the gravity term in Equation~(\ref{momentum-total}) is replaced by
$-2q\Omega^2x\hat{\bf e}_x$
(i.e., put ${\bf\Omega}\times({\bf\Omega}\times{\bf r})-{\bf g}=-2q\Omega^2x\hat{\bf e}_x$).
This is the tidal expansion of the effective potential \citep[see e.g.,][]{hawley1995}.

\subsubsection{Initial equilibrium state of shearing box model}\label{sec:equil-shear1}
We adopt the following state as the initial equilibrium state of the shearing box model.
Density, plasma pressure, CR pressure, magnetic field strength are taken as constant.
The components of the magnetic field and the velocity are chosen as
\begin{equation}\label{icshear1}
  B_{x}=B_{y}=V_{x}=V_{z}=0\,,
\end{equation}
\begin{equation}\label{icshear2}
  V_{y}=-q{\Omega}x\,,
\end{equation}
\begin{equation}\label{ishear3}
  B_z=\left({2\mu_0 P_{\rm g}\over\beta}\right)^{1/2}\,,
\end{equation}
where $\beta$ is the initial ratio of the magnetic pressure to the thermal plasma pressure.
We set the initial CR pressure as $P_{\rm c}=\alpha P_{\rm g}$ and
$P_{\rm g}=\rho_0 C_{{\rm s}0}^2/\gamma_{\rm g}$ is the thermal plasma pressure (and $C_{{\rm s}0}$ is the sound speed).
Setting $q=3/2$ in Equation~(\ref{icshear2}) gives the Keplerian rotation.
We set the units of calculation as follows:
the units of density, velocity and length are $\rho_0=1.6\times 10^{-24}$ g cm$^{-3}$,
$C_{{\rm s}0}=10^6$ cm s$^{-1}$ and $H_0=3\times 10^{20}$ cm, respectively.
Figure~\ref{fig2} shows the distribution of the normalized physical values stated above.
In this example, we take $\gamma_{\rm g}=5/3$, $\gamma_{\rm c}=4/3$ and $\alpha=1$.

\subsection{Differentially rotating cylinder}\label{sec:cyl1}
\begin{figure}
\epsscale{.80}
\plotone{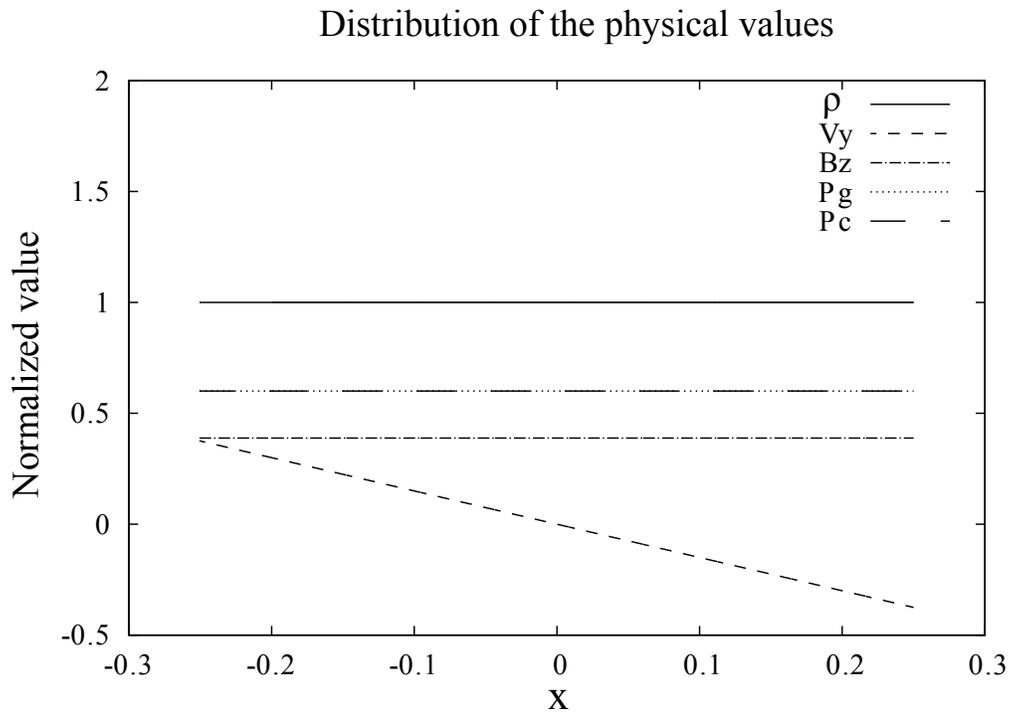}
\caption{
Initial distribution of the physical quantities in the shearing box model.
}
\label{fig3}
\end{figure}

\begin{figure}
\epsscale{.80}
\plotone{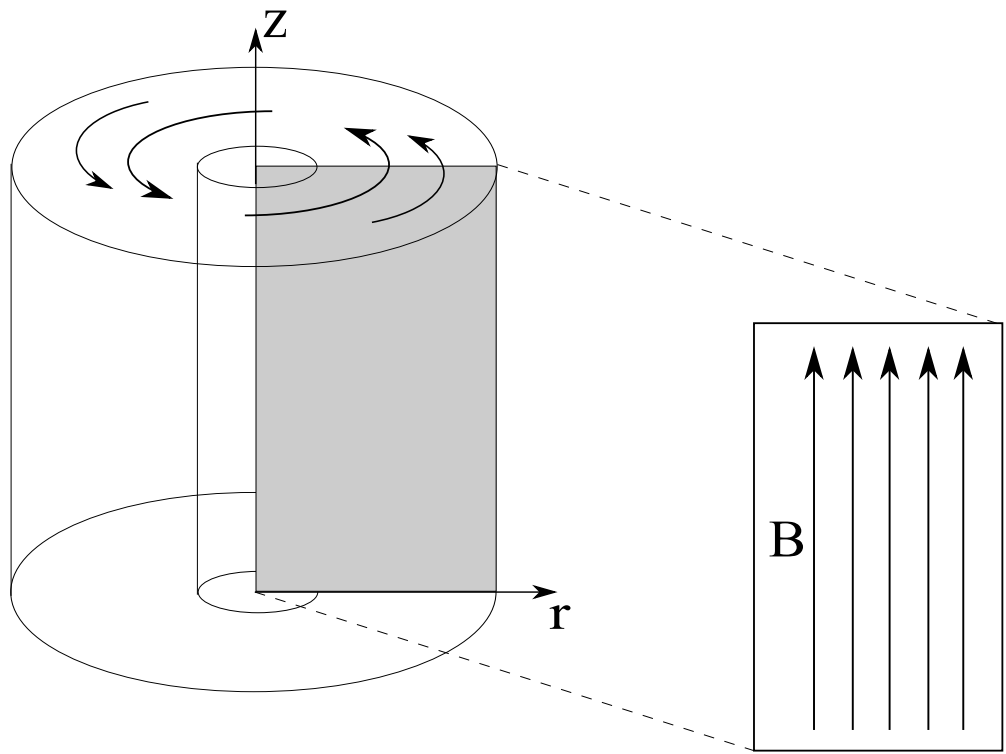}
\caption{Schematic picture of the differentially rotating cylinder model.}
\label{fig2}
\end{figure}

Another model of interest is the differentially rotating cylinder model.
In this model we use the cylindrical coordinate $(r,\phi,z)$
and consider the system in inertial frame, i.e., put ${\bf\Omega}=0$
in Equation~(\ref{momentum-total}) but keeping the gravity term
(see Figure~\ref{fig3}).

\subsubsection{Initial equilibrium state of differentially rotating cylinder model}\label{sec:equil-cyl1}
We adopt the following state as the initial equilibrium in the case of differentially rotating cylinder model.
The equilibrium distribution of a rotating cylinder is obtained from the Newtonian analogue
of the relativistic tori of \citet{abramowicz}.
Since we are interested in regions close to the equatorial plane, i.e., $z\ll r$, hence for simplicity
we assume that the initial equilibrium state depends on $r$ only, and
\begin{equation}\label{iccyl1}
  {\bf V}=V_\phi\hat{\bf e}_\phi+V_z\hat{\bf e}_z\,,
  \quad {\bf B}=B_\phi\hat{\bf e}_\phi+B_z\hat{\bf e}_z\,.
\end{equation}
We note that in this case the diffusion term in Equation~(\ref{cr-energy}) vanishes.
Momentum balance in $\hat{\bf e}_r$ gives
\begin{equation}\label{iccyl2}
  {1\over\rho}{\!d\over d r} \left[P_{\rm g}+P_{\rm c}+{(B_\phi^2+B_z^2)\over 2\mu_0}\right]
  -\,{1\over r}\left(V_\phi^2-{B_\phi^2\over\mu_0\rho}\right)-g_r=0\,.
\end{equation}
To illustrate ideas, we take the initial total pressure (sum of thermal pressure and CR pressure) in the rotating torus as,
\begin{equation}\label{iccyl3}
P_{\rm g} + P_{\rm c} = P_{\rm sum} = K\rho^{1+1/n}\,,
\quad
P_{\rm c} = \alpha P_{\rm g}\,.
\end{equation}
Note that a change in $\alpha$ does not change the density distribution.
This is more convenient when we analyze the dependence of the MRI growth rate on $\alpha$.
We assume $B_z=B_0$ is constant, $B_\phi=0$, ${\bf g}=-{\bf\nabla}\Psi$,
and the distribution of specific angular momentum ($L=r V_\phi$) as
\begin{equation}\label{iccyl4}
  L=L_0\left({r\over r_0}\right)^a\,,
\end{equation}
then the density distribution of the rotating plasma torus is determined by
\begin{equation}\label{iccyl5}
  (n+1){(1+\alpha)P_{\rm g}\over\rho}
  -\,{L_0^2\over 2(a-1)r_0^2}\left(r\over r_0\right)^{2(a-1)}+\Psi
  ={\cal E}\,,
\end{equation}
where ${\cal E}$ is a constant (cf. Bernoulli theorem in fluid physics).
In the rest of the paper, we consider the gravitational potential is dominated by a point mass at the center,
$\Psi=- GM/\sqrt{r^2+z^2}\approx -GM/r$.
We consider a non-rotating high-temperature halo outside the rotating plasma torus.
We take isothermal equation of state for the halo, and adopt the distribution
\begin{equation}\label{iccyl6}
  \rho=\rho_{\rm h} \exp\left[{1\over\epsilon_{\rm h}}\left({r_0\over r}-1\right)\right]\,,
\end{equation}
where $\rho_{\rm h}$ is the density of the halo at $r=r_0$.
Here $\epsilon_{\rm h}=C_{\rm sh}^2/V_{{\rm K}0}^2$,
where $C_{\rm sh}$ and $V_{{\rm K}0}$ are the isothermal sound speed (in the halo)
and the Keplerian velocity at $r=r_0$.
We take $r_0$ as the radius at which the density of the rotating plasma torus is maximum,
and this density is denoted as $\rho_0$.
We set the units of length, velocity, time and density as $r_0$, $V_{{\rm K}0}$,
$r_0/V_{{\rm K}0}$ and $\rho_0$, respectively.
Subsequently, we have two nondimensional parameters
for the initial torus
\begin{equation}\label{iccyl7}
  \epsilon_{\rm th}={C_{{\rm s}0}^2\over\gamma_{\rm g} V_{{\rm K}0}^2}\,,
  \quad \epsilon_{\rm B}={V_{{\rm A}0}^2\over V_{{\rm K}0}^2}\,,
\end{equation}
where $C_{{\rm s}0}=(\gamma_{\rm g} P_{{\rm g}0}/\rho_0)^{1/2}$
is the sound speed in the torus at $r=r_0$,
and $V_{{\rm A}0}=(B_0^2/\mu_0\rho_0)^{1/2}$ the Alfv\'en speed at $r=r_0$.
$\gamma_{\rm g}$ is the adiabatic index of the thermal plasma in the torus.
In fact,
if we represent the gravitational energy by $\rho V_{{\rm K}0}^2/2$, then
$\epsilon_{\rm B}$ is the ratio of magnetic energy to gravitational energy at $r=r_0$, and
$\epsilon_{\rm th}$ is $(\gamma_{\rm g}-1)/2$ times the ratio of thermal energy to gravitational energy at $r=r_0$.

\begin{figure}
\epsscale{.80}
\plotone{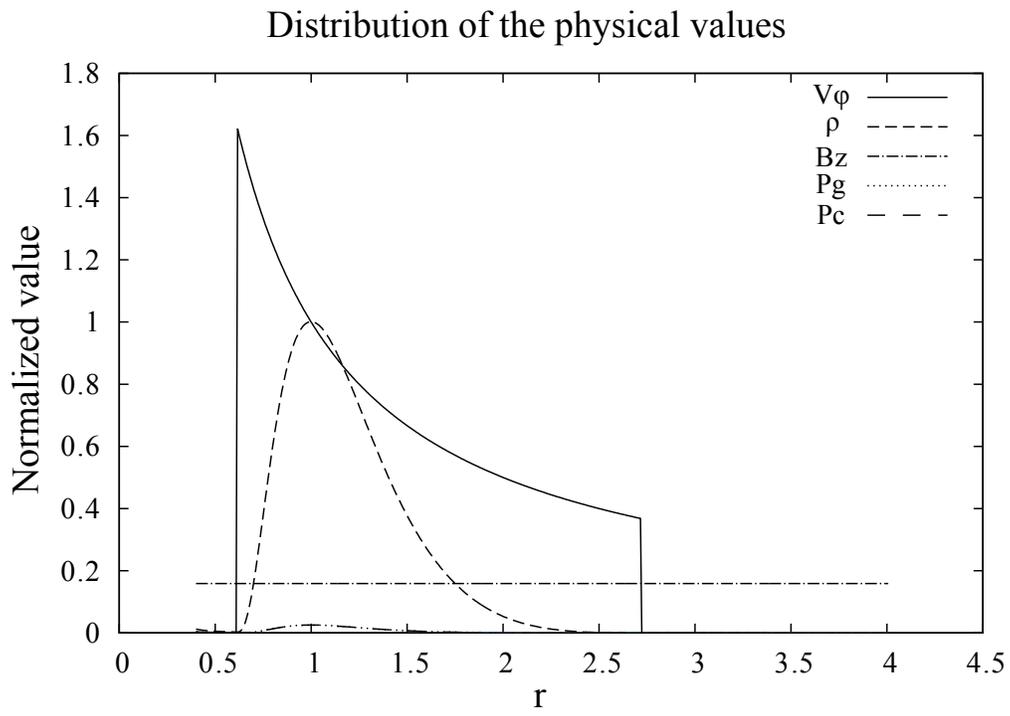}
\caption{
Initial distribution of the physical quantities in the differentially rotating cylinder model.
}
\label{fig4}
\end{figure}

As an example, we pick $n=3$, $a=0$ (i.e., $L$ is constant), $\rho_{\rm h}/\rho_0=10^{-3}$,
$\epsilon_{\rm h}=1.0$,
$\epsilon_{\rm th}=5.0\times10^{-2}$, $\epsilon_{\rm B}=4.0\times10^{-4}$.
Figure~\ref{fig4} shows the distribution of the normalized physical values stated above.
In this example, we take $\gamma_{\rm g}=5/3$, $\gamma_{\rm c}=4/3$ and
$\alpha = 1$.
The equilibrium model presented here is a modification of the one in \citet{kuwabara2005} to include CRs.

\section{LINEAR STABILITY ANALYSIS}\label{sec:linear}

We perform standard linear stability analysis on the set of equations~(\ref{mass-total})--(\ref{mag-energy}).
Recall that in the shearing box model the term
${\bf\Omega}\times({\bf\Omega}\times{\bf r})-{\bf g}=-2q\Omega^2x\hat{\bf e}_x$,
while in the differentially rotating cylinder model ${\bf\Omega}=0$ and ${\bf g}=-{\bf\nabla}\Psi$.

In the following analysis, the unperturbed background we consider depends only on one coordinate
and the velocity and magnetic field is orthogonal to this coordinate axis
(this is slightly more general than the initial equilibrium state described in previous section).

\subsection{Shearing box}\label{sec:shear2}
In the shearing box model, we denote the set of physical quantities of interest as
$\chi=\{\rho,V_x,V_y,V_z,P_{\rm g},P_{\rm c},B_x,B_y,B_z\}$ and the perturbed quantities
$\delta\chi=\{\delta\rho,\delta V_x,\delta V_y,\delta V_z,\delta P_{\rm g},
\delta P_{\rm c},\delta B_x,\delta B_y,\delta B_z\}$.
We consider the perturbation of the form
\begin{equation}\label{pert-shear}
  \delta\chi(t,x,y,z)=\delta{\bar\chi}(x)\exp\left(\sigma t + i\,k_y y+i\,k_z z \right)\,,
\end{equation}
where $\delta\bar{\chi}=\{\delta\bar{\rho},\delta\bar{V_x},i\,\delta\bar{V_y},i\,\delta\bar{V_z},\delta\bar{P_{\rm g}},
\delta\bar{P_{\rm c}},-i\,\delta\bar{B_x},\delta\bar{B_y},\delta\bar{B_z}\}$.
After some manipulations, the set of linear perturbation equations can be reduced to two first order ODEs.
In fact, these two ODEs are the continuity equation and the $x$-momentum equation.
Explicitly,
\begin{eqnarray}\label{pert-eq-shear}
  {\! d\over d x}\left[\begin{array}{c}
                         \delta\bar{V_x} \\
                         \delta\bar{P_{\rm t}}
                       \end{array}\right]
                =\left[\begin{array}{cc}
                         A_{11} & A_{12} \\
                         A_{21} & A_{22}
                       \end{array}\right]
                 \left[\begin{array}{c}
                         \delta\bar{V_x} \\
                         \delta\bar{P_{\rm t}}
                       \end{array}\right]\,,
\end{eqnarray}
where
\begin{equation}\label{dPt-shear}
  \delta\bar{P_{\rm t}} = \delta\bar{P_{\rm g}} + \delta\bar{P_{\rm c}}
  + {1\over \mu_0}\left({B_y\delta\bar{B_y}}+{B_z\delta\bar{B_z}}\right)\,,
\end{equation}
\begin{eqnarray}\label{A11-shear}
  A_{11} &=& {1\over(1+W){\cal A}^2}
  \left[-\,{1\over\rho}{dP_{\rm t}\over dx}
  -i\,{2\Omega V_{\rm A}^2B_y\left(k_yB_y + k_zB_z \right)\over(1+W)\Sigma \left(B_y^2+B_z^2\right)} \right] \nonumber \\
  &&\quad + i\,{2\Omega k_y\over(1+W)\Sigma} + {i\over\Sigma}{\! d\over dx}\left(k_yV_y+k_zV_z \right)\,,
\end{eqnarray}
\begin{eqnarray}\label{A12-shear}
  A_{12} = -\,{\Sigma\over\rho(1+W)^2{\cal A}^2}
  -\,{\left(k_y^2+k_z^2\right)\over\rho (1+W)\Sigma}\,,
\end{eqnarray}
\begin{eqnarray}\label{A21-shear}
  A_{21} &=& -\,{\rho\over\Sigma}\left\{ (1+W)\Sigma^2 + 2\Omega{dV_y\over dx}+{4\Omega^2\over(1+W)}
  +{1\over\rho^2}{d\rho\over dx}{dP_{\rm t}\over dx} \right. \nonumber \\
  &&\quad\left. -\,{1\over{\cal A}^2}\left[-\,{1\over\rho}{dP_{\rm t}\over dx}
  -i\,{2\Omega V_{\rm A}^2B_y(k_yB_y+k_zB_z)\over(1+W)\Sigma \left(B_y^2+B_z^2\right)} \right]^2 \right\}\,,
\end{eqnarray}
\begin{eqnarray}\label{A22-shear}
  A_{22} = -\,{1\over(1+W){\cal A}^2}\left[-\,{1\over\rho}{dP_{\rm t}\over dx}
  -i\,{2\Omega V_{\rm A}^2B_y(k_yB_y+k_zB_z)\over(1+W)\Sigma \left(B_y^2+B_z^2\right)}\right]
  -i\,{2\Omega k_y\over(1+W)\Sigma}\,,
\end{eqnarray}
and
\begin{equation}\label{Sigma-shear}
  \Sigma = \sigma + i\,k_yV_y + i\,k_zV_z\,,
\end{equation}
\begin{equation}\label{A-shear}
  {\cal A}^2 = C_{\rm s}^2 + {C_{\rm c}^2\over(1+K)} + {V_{\rm A}^2\over(1+W)}\,,
\end{equation}
\begin{equation}\label{W-shear}
  W = {V_{\rm A}^2\left(k_yB_y + k_zB_z \right)^2\over\Sigma^2 \left(B_y^2+B_z^2\right)}\,,
\end{equation}
\begin{equation}\label{K-shear}
  K = {\kappa_{\|}\left(k_yB_y + k_zB_z \right)^2\over\Sigma \left(B_y^2+B_z^2\right)}\,,
\end{equation}
\begin{equation}\label{speed-shear}
  C_{\rm s}^2={\gamma_{\rm g}P_{\rm g}\over\rho}\,,
  \quad C_{\rm c}^2={\gamma_{\rm c}P_{\rm c}\over\rho}\,,
  \quad V_{\rm A}^2={\left(B_y^2+B_z^2\right)\over\mu_0\rho}\,,
\end{equation}
\begin{equation}\label{Pt-shear}
  P_{\rm t} = P_{\rm g} + P_{\rm c} + {\left(B_y^2+B_z^2\right)\over 2\mu_0}\,.
\end{equation}
The other perturbed quantities can be expressed algebraically in terms of $\delta\bar{V_x}$ and $\delta\bar{P_{\rm t}}$
(see Appendix~A).

\subsubsection{Result of shearing box model}\label{sec:result-shear2}
We take the initial equilibrium state described in section~\ref{sec:equil-shear1} as the unperturbed state.
the boundary conditions at $x=0.25$ in Figure~\ref{fig2} are taken as
$\delta\bar{V_x}=1+i\,0$ and $\delta\bar{P_{\rm t}}=0+i\,0$.
This condition allows perturbation of the flow to pass through the boundary in the $x$-direction.
Moreover, the total pressure is held constant on this boundary.
On the other boundary at $x=-0.25$, we require $\Re(\delta\bar{V_x})\ne 0$ and $\Re(\delta\bar{P_{\rm t}})=0$.
(This carries the same meaning as the conditions at $x=0.25$.)

We solve the set of linearized perturbation equations, the set of ODEs~(\ref{pert-eq-shear}) by shooting method.
For a trial value of $\sigma$, we integrate each equation from the boundary at $x=0.25$ (with the assigned boundary value)
to the boundary at $x=-0.25$.
We then adjust the value of $\sigma$ until $\delta\bar{P_{\rm t}}$ matches the boundary condition at $x=-0.25$.
We take this value of $\sigma$ as the eigenvalue,
and take the maximum value of $\sigma$ as the maximum growth rate of the system.

\begin{figure}
\epsscale{1.0}
\plotone{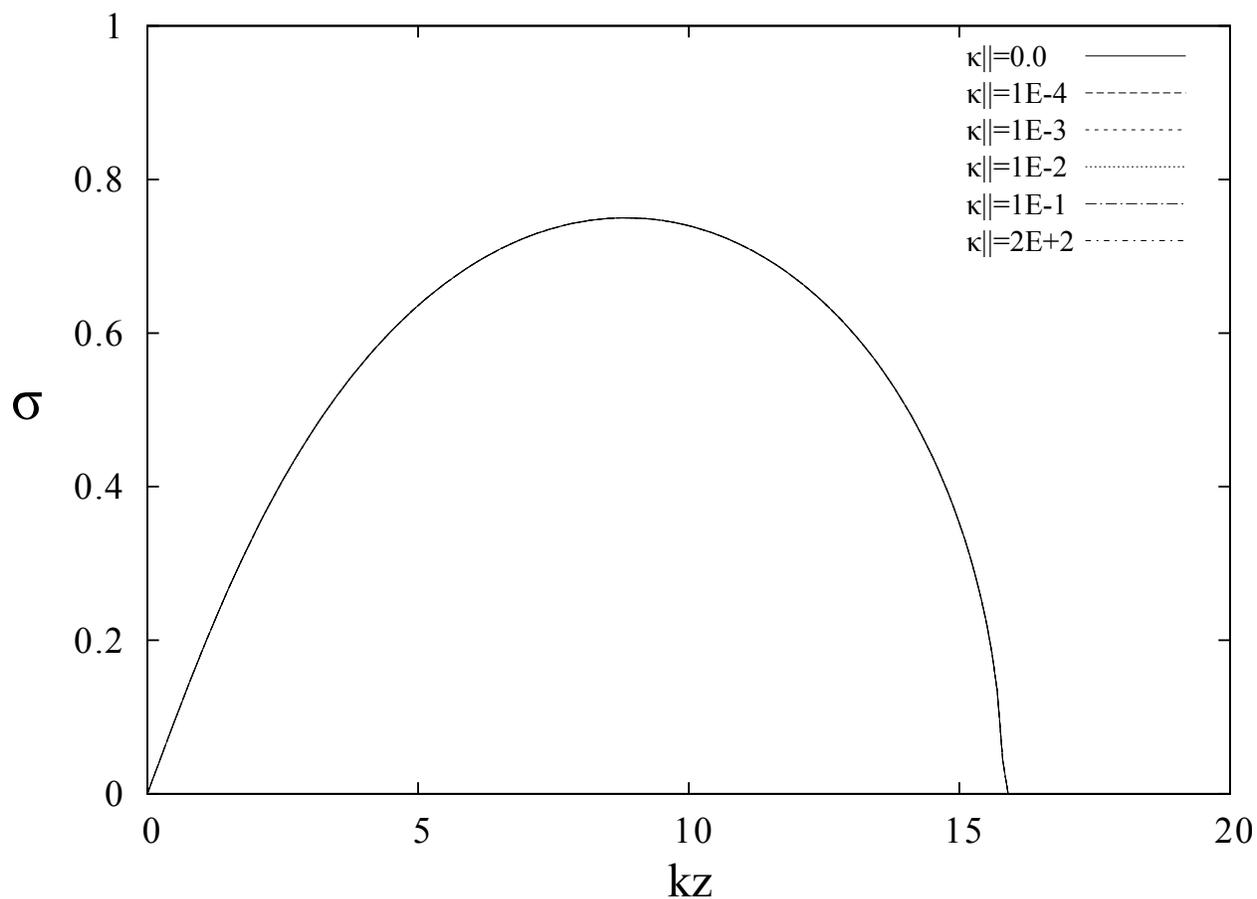}
\caption{
Dispersion relation for the magnetorotational instablity with the effect of CRs for different
$\kappa_{\|}$ in the shearing box model.
Here $\sigma$ is the growth rate of perturbation and $k_z$ is the
wavenumber along the direction of the magnetic field in the unperturbed state.
Apparently, all the cases collapse to one line. This indicates the dispersion relation is
almost independent of $\kappa_{\|}$ in the shearing box model. The reason is CR pressure
is uniform in the initial unperturbed state.
}
\label{fig5}
\end{figure}

\begin{figure}
\epsscale{1.0}
%
%
\plotone{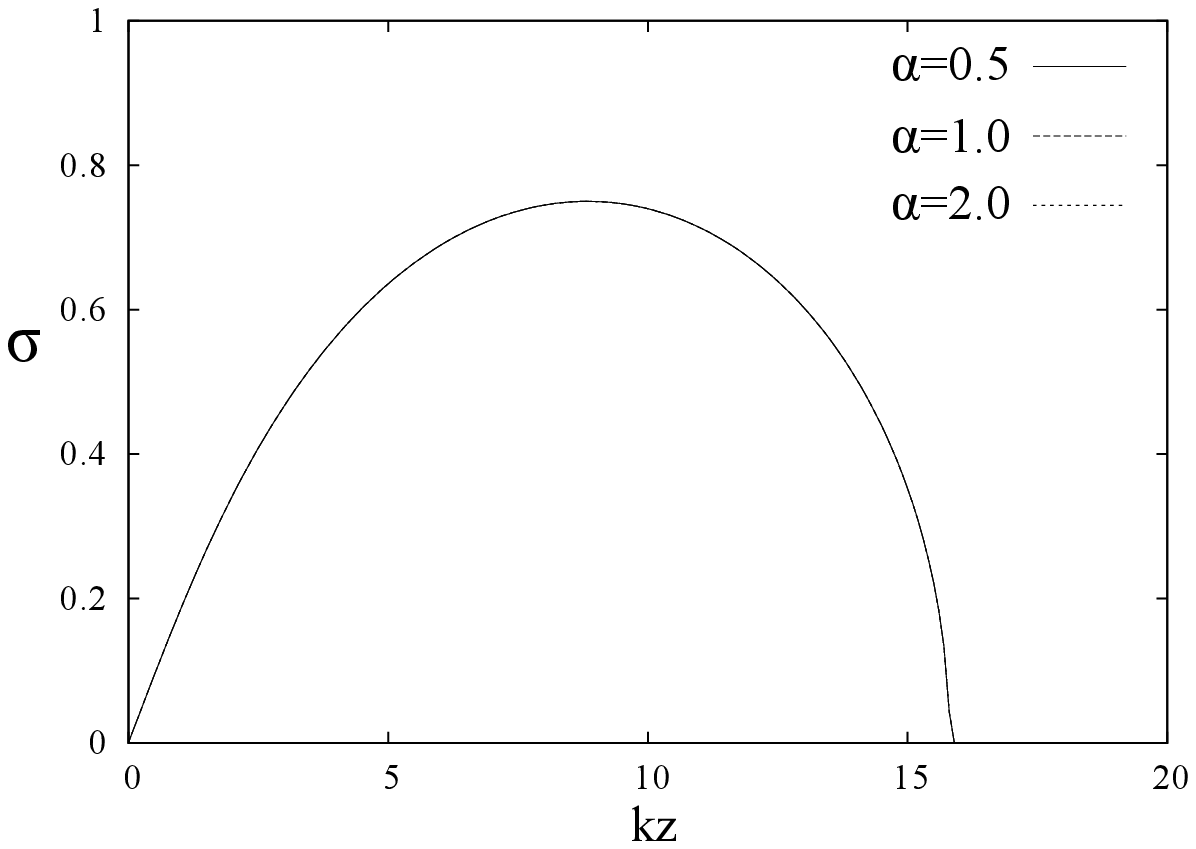}
\caption{
Same as Figure~\ref{fig5} except that the CR diffusion coefficient is fixed at $\kappa_{\|}=200$ and
the ratio of CR pressure to thermal pressure $\alpha$ varies (while the sum of them is kept fixed).
}
\label{fig6}
\end{figure}

Figure~\ref{fig5} shows the result of the linear stability analysis of the shearing box model.
The figure displays the dispersion relation for different CR diffusion coefficient $\kappa_{\|}$.
In the figure, $\sigma$ is the growth rate, and $k_z$ is the wave number in the the direction of the initial magnetic field.
Here we take the CR diffusion coefficient as an input parameter
and other quantities as fixed parameters (e.g., the ratio of the CR pressure to the gas pressure $\alpha = 1$,
the ratio of the gas pressure to the magnetic pressure $\beta = 100$, the rotational angular frequency $\Omega = 1$).
The maximum value of the normalized $\kappa_{\|}=200$ in Figure~\ref{fig5} corresponds to
$\kappa_{\|}=3\times10^{28}$ cm$^2$ s$^{-1}$, the value estimated in our Galaxy \citep{berezinskii, ptuskin, ryu}.
The maximum growth rate is given at $k_z\sim 8.8$ and the cut-off wave number where the growth rate
becomes zero is $k_z\sim 15.9$.
In Figure~\ref{fig5}, The dispersion relations for different $\kappa_{\|}$ almost completely overlap each other,
therefore, we can see only one curve in this scale.
Figure~\ref{fig6} shows the dispersion relation for different $\alpha$.
In this figure, the value of $\kappa_{\|}=200$ is fixed and the other parameters are the same as in Figure~\ref{fig5}.
The dispersion relations for different $\alpha$ also almost completely overlap each other.
We point out that the two profiles of Figures~\ref{fig5} \& \ref{fig6} are the same.
In this model, neither the ratio of CR pressure to thermal pressure (while the sum is kept constant) nor the diffusion of CR
will affect the growth rate significantly.

\subsection{Differentially rotating cylinder}\label{sec:cyl2}
In the differentially rotating cylinder model, we denote the set of physical quantities of interest as
$\chi^\prime=\{\rho,V_r,V_\phi,V_z,P_{\rm g},P_{\rm c},B_r,B_\phi,B_z\}$ and the perturbed quantities
$\delta\chi^\prime=\{\delta\rho,\delta V_r,\delta V_\phi,\delta V_z,\delta P_{\rm g},
\delta P_{\rm c},\delta B_r,\delta B_\phi,\delta B_z\}$.
We consider the perturbation of the form
\begin{equation}\label{pert-cyl}
  \delta\chi^\prime(t,x,y,z)=\delta{\bar\chi}^\prime(x)\exp\left(\sigma t + i\,m\phi+i\,k_z z \right)\,,
\end{equation}
where $\delta\bar{\chi}^\prime=\{\delta\bar{\rho},\delta\bar{V_r},i\,\delta\bar{V_\phi},i\,\delta\bar{V_z},\delta\bar{P_{\rm g}},
\delta\bar{P_{\rm c}},-i\,\delta\bar{B_r},\delta\bar{B_\phi},\delta\bar{B_z}\}$.
Again the set of linear perturbation equations can be reduced to two first order ODEs,
and these two ODEs are the continuity equation and the $r$-momentum equation.
Explicitly,
\begin{eqnarray}\label{pert-eq-cyl}
  {\! d\over d r}\left[\begin{array}{c}
                         \delta\bar{V_r} \\
                         \delta\bar{P_{\rm t}}^\prime
                       \end{array}\right]
                =\left[\begin{array}{cc}
                         A^\prime_{11} & A^\prime_{12} \\
                         A^\prime_{21} & A^\prime_{22}
                       \end{array}\right]
                 \left[\begin{array}{c}
                         \delta\bar{V_r} \\
                         \delta\bar{P_{\rm t}}^\prime
                       \end{array}\right]\,,
\end{eqnarray}
where
\begin{equation}\label{dPt-cyl}
  \delta\bar{P_{\rm t}}^\prime = \delta\bar{P_{\rm g}} + \delta\bar{P_{\rm c}}
  + {1\over \mu_0}\left({B_\phi\delta\bar{B_\phi}}+{B_z\delta\bar{B_z}}\right)\,,
\end{equation}
\begin{eqnarray}\label{A11-cyl}
  A^\prime_{11}&=&{1\over(1+W^\prime){\cal A^\prime}^2}
  \left[-{1\over\rho}{d P^\prime_{\rm t}\over dr}
  +{(1-W^\prime){V^\prime_{\rm A}}^2 B_\phi^2\over(1+W^\prime)r\left(B_\phi^2+B_z^2\right)} \right. \nonumber \\
  &&\quad \left. -i\,{2\Omega {V^\prime_{\rm A}}^2 B_\phi\over(1+W^\prime)\Sigma^\prime \left(B_\phi^2+B_z^2\right)}
  \left({m\over r}B_\phi+k_z B_z \right)  \right] \nonumber \\
  &&\quad+{2{V^\prime_{\rm A}}^2 m B_{\phi}\over(1+W^\prime){\Sigma^\prime}^2 r^2\left(B_\phi^2+B_z^2\right)}
  \left({m\over r}B_\phi+k_zB_z \right)  \nonumber \\
  &&\quad +i\,{2m\Omega\over(1+W^\prime)\Sigma^\prime r}-{1\over r}
  +{1\over\Sigma^\prime}{d\Sigma^\prime\over dr}\,,
\end{eqnarray}
\begin{eqnarray}\label{A12-cyl}
  A^\prime_{12}=-\,{\Sigma^\prime\over\rho(1+W^\prime)^2{\cal A^\prime}^2}-{1\over\rho(1+W^\prime)\Sigma^\prime}
  \left({m^2\over r^2}+k_z^2 \right)\,,
\end{eqnarray}
\begin{eqnarray}\label{A21-cyl}
  A^\prime_{21}&=&-\,{\rho\over\Sigma^\prime}\left\{(1+W^\prime){\Sigma^\prime}^2
  +2r\Omega{d\Omega\over dr}+{4\Omega^2\over(1+W^\prime)} \right. \nonumber \\
  &&\quad + \left. {4\over r}\left[{(1-W^\prime){V^\prime_{\rm A}}^2 B_\phi^2\over(1+W^\prime)r\left(B_\phi^2+B_z^2\right)}
  -i\,{2\Omega {V^\prime_{\rm A}}^2 B_\phi\over(1+W^\prime)\Sigma^\prime\left(B_\phi^2+B_z^2\right)}
  \left({m\over r}B_\phi+k_z B_z \right) \right] \right. \nonumber \\
  &&\quad + \left. {{V^\prime_{\rm A}}^2B_{\phi}^2\over r\left(B_\phi^2+B_z^2\right)}
  \left[-\,{2\over B_\phi}{dB_\phi\over dr}+{1\over\rho}{d\rho\over dr}
  -\,{2(1-W^\prime)\over r(1+W^\prime)} \right] +{1\over\rho^2}{d\rho\over dr}{d P^\prime_{\rm t}\over dr} \right. \nonumber \\
  &&\quad \left. - \, {1\over {\cal A^\prime}^2}\left[-\,{1\over\rho}{d P^\prime_{\rm t}\over dr}
  +{(1-W^\prime){V^\prime_{\rm A}}^2 B_\phi^2\over(1+W^\prime)r\left(B_\phi^2+B_z^2\right)} \right. \right. \nonumber \\
  &&\quad \left.\left.  -i\,{2\Omega {V^\prime_{\rm A}}^2 B_\phi\over(1+W^\prime)
  \Sigma^\prime\left(B_\phi^2+B_z^2\right)}\left({m\over r}B_\phi+k_z B_z \right) \right]^2 \right\}\,,
\end{eqnarray}
\begin{eqnarray}\label{A22-cyl}
  A^\prime_{22}&=&-\,{1\over(1+W^\prime){\cal A^\prime}^2}\left[-\,{1\over\rho}{{d P^\prime_{\rm t}}\over dr}
  +{(1-W^\prime){V^\prime_{\rm A}}^2 B_\phi^2\over(1+W^\prime)r\left(B_\phi^2+B_z^2\right)} \right. \nonumber \\
  &&\quad \left. -i\,{2\Omega {V^\prime_{\rm A}}^2 B_\phi\over(1+W^\prime)\Sigma^\prime\left(B_\phi^2+B_z^2\right)}
  \left({m\over r}B_\phi+k_z B_z\right) \right. \nonumber \\
  &&\quad + \left. {2{V^\prime_{\rm A}}^2 m B_\phi\over(1+W^\prime){\Sigma^\prime}^2 r^2\left(B_\phi^2+B_z^2\right)}
  \left({m\over r}B_\phi+k_z B_z\right)+i\,{2m\Omega\over(1+W^\prime)\Sigma^\prime r} \right]\,,
\end{eqnarray}
and
\begin{equation}\label{Sigma-cyl}
  \Sigma^\prime = \sigma + i\,m \Omega + i\,k_zV_z\,,
\end{equation}
\begin{equation}\label{A-cyl}
  {\cal A^\prime}^2 = C_{\rm s}^2 + {C_{\rm c}^2\over(1+K^\prime)} + {{V^\prime_{\rm A}}^2\over(1+W^\prime)}\,,
\end{equation}
\begin{equation}\label{W-cyl}
  W^\prime = {{V^\prime_{\rm A}}^2\over{\Sigma^\prime}^2 \left(B_\phi^2+B_z^2\right)}\left({m\over r}B_\phi + k_zB_z \right)^2\,,
\end{equation}
\begin{equation}\label{K-cyl}
  K^\prime = {\kappa_{\|}\over\Sigma^\prime \left(B_\phi^2+B_z^2\right)}\left({m\over r}B_\phi + k_zB_z \right)^2\,,
\end{equation}
\begin{equation}\label{speed-cyl}
  C_{\rm s}^2={\gamma_{\rm g}P_{\rm g}\over\rho}\,,
  \quad C_{\rm c}^2={\gamma_{\rm c}P_{\rm c}\over\rho}\,,
  \quad {V^\prime_{\rm A}}^2={\left(B_\phi^2+B_z^2\right)\over\mu_0\rho}\,,
\end{equation}
\begin{equation}\label{Pt-cyl}
  P^\prime_{\rm t} = P_{\rm g} + P_{\rm c} + {\left(B_\phi^2+B_z^2\right)\over 2\mu_0}\,,
\end{equation}
\begin{equation}\label{Omega-cyl}
  \Omega={V_\phi\over r}\,.
\end{equation}
The other perturbed quantities can be expressed algebraically in terms of $\delta\bar{V_r}$ and $\delta\bar{P_{\rm t}}^\prime$
(see Appendix~A).

\subsubsection{Result of Differentially rotating cylinder model}\label{sec:result-cyl2}
We take the initial equilibrium state described in section~\ref{sec:equil-cyl1} as the unperturbed state.
Similar to the shearing box model, the boundary conditions at the outer boundary $r=4.0$ in Figure~\ref{fig4} are taken as
$\delta\bar{V_r}=1+i\,0$ and $\delta\bar{P_{\rm t}}^\prime=0+i\,0$.
Hence perturbation of the flow can pass through the boundary in the $r$-direction.
Moreover, the total pressure is held constant on the outer boundary.
At the inner boundary $r=0.4$, we also require $\Re(\delta\bar{V_r})\ne 0$ and $\Re(\delta\bar{P_{\rm t}}^\prime)=0$.

Similar to the case of the shearing box model, we solve the set of linearized perturbation equations of the differentially
rotating cylinder model Equation~\ref{pert-eq-cyl}) by shooting method.
For a trial value of $\sigma$, we integrate each equation from the boundary at $r=4.0$ (with the assigned boundary value)
to the boundary at $r=0.4$.
We then adjust the value of $\sigma$ until $\delta\bar{P_{\rm t}}^\prime$ matches the boundary condition at $r=0.4$.
We take this value of $\sigma$ as the eigenvalue,
and take the maximum value of $\sigma$ as the maximum growth rate of the system.

Figure~\ref{fig7} shows the result of the linear stability analysis of the differentially rotating cylinder model.
The left panel of the figure displays the dispersion relation for different CR diffusion coefficient $\kappa_{\|}$.
Here $\kappa_{\|}=0.4$ corresponds to the nominal value in our Galaxy $\kappa_{\|}=3\times10^{28}$ cm$^2$ s$^{-1}$.
The cut-off wave number where the growth rate becomes zero takes the same value for different values of $\kappa_{\|}$
except when $\kappa_{\|}=0.0$.
In the case of $\kappa_{\|}=0.0$, the cut-off wave number is about $5.4\%$ smaller.
This can be traced back to the fact that the unstable mode of the non-zero $\kappa_{\|}$ case (between the two cut-off wavenumbers)
becomes neutrally stable (growth rate equals zero) when $\kappa_{\|}$ turns to zero exactly.
The cut-off wave number in the case of $\kappa_{\|}=0.0$ is smaller because the unstable criterion depends on the combine
pressures of plasma and CRs (compare to plasma pressure only in the case of $\kappa_{\|}>0.0$ as CR diffuse through the plasma).
Similar result was obtained in \citet{kuwabara2006} for the role of CRs on Parker-Jeans instability.
The maximum growth rate becomes larger as $\kappa_{\|}$ increases.
The right panel of Figure~\ref{fig7} shows the dependence of the maximum growth rate on $\kappa_{\|}$.
Note the horizontal axis is in log scale.
The maximum growth rate does not change much when $\kappa_{\|}<0.0005$,
then it increases considerably in the range $0.0005\le\kappa_{\|}\le 0.05$,
and then kind of saturated when $\kappa_{\|}>0.05$.

Figure~\ref{fig8} shows the growth rate dependence on $\alpha$, the ratio of CR pressure to thermal plasma pressure.
In this figure, the diffusion coefficient is fixed at $\kappa_{\|}=200$.
The larger is $\alpha$ the larger is the growth rate and the larger the cut-off wavenumber.

\section{2.5-DIMENSIONAL SIMULATION}\label{sec:simulation}

In this section, we solve the MHD equations combined with the CR energy equation,
Equations~(\ref{mass-total})--(\ref{mag-energy}), by MHD simulation code augmented with CR.
For the shearing box model we put the term
${\bf\Omega}\times({\bf\Omega}\times{\bf r})-{\bf g}=-2q\Omega^2x\hat{\bf e}_x$,
and for the differentially rotating cylinder model we set ${\bf\Omega}=0$ and ${\bf g}=-{\bf\nabla}\Psi$.
The MHD simulations are 2.5-dimensional nonlinear, time-dependent, and compressible in cartesian coordinate
for the shearing box model, and in cylindrical coordinate
for the differentially rotating cylinder model.
In \citet{kuwabara2004}, we used a hybrid scheme to simulate the CR-MHD system.
We used the Lax-Wendroff scheme for the MHD part and the biconjugate gradients stabilized
(BiCGstab) method for the diffusion part of the CR energy equation as described in \citet{yokoyama2001}
to reduce computation time.
However, in this work we use the Lax-Wendroff scheme for all the equations (MHD and CR equations),
because computer is very powerful nowadays.
The calculation time for such 2.5-dimensional simulation is rather short.

We adopt the MHD code developed by \citet{shibata1983} and subsequently extended by \citet{matsumoto1996, hayashi1996}.
Currently, this MHD code is incorporated in the Coordinated Astronomical Numerical Software
(CANS)\footnote{http://www.astro.phys.s.chiba-u.ac.jp/cans} and anyone can use it under
the acceptance of their licenses.

\subsection{Numerical results of the shearing box model}
In the shearing box model, we calculate within the region extracted from $x$-$z$ plane as shown in Figure~\ref{fig1}.
The size of this region is $0.5H_0\times 1.0H_0$, with $x\in[-0.25H_0,0.25H_0]$ and $z\in[0.0H_0,1.0H_0]$.
The numerical grid resolution and the grid size are
$N_x=41$, $N_z=82$, and $\Delta x=0.0125H_0$, $\Delta z=0.0125H_0$.
We assume a periodic boundary at $x=x_{\rm min}$, $x=x_{\rm max}$, and at $z=z_{\rm min}$, $z=z_{\rm max}$.
The initial equilibrium state is described in section~\ref{sec:equil-shear1}.
To start the simulation, a small velocity perturbation is added to the initial equilibrium as follows,
\begin{equation}\label{Vy-sim-shear}
  \delta V_y = 10^{-3}\times\sin(k_z z)\,.
\end{equation}
We choose $k_z=10$ as a reference to the result of linear analysis (see Figure~\ref{fig5}).

We study two values of the CR diffusion coefficient, $\kappa_{\|}=10^{-4}$, and $\kappa_{\|}=10.0$
as the representative values (see the right panel of Figure~\ref{fig7}).
We should point out that Figure~\ref{fig7} is the result of linear stability analysis of the differentially rotating cylinder model.
The maximum growth rate $\sigma_{\rm max}$ is low for $\kappa_{\|}=10^{-4}$, while it is high for $\kappa_{\|}=10.0$.
In fact, the linear analysis on the shearing box model showed that the growth rate is almost the same for different
$\kappa_{\|}$ (see Figure~\ref{fig5}). This is confirmed by MHD simulations (see below).

Figure \ref{fig9} shows the time evolution of the distributions of the magnetic field and the CR pressure.
In the figure the white curves are the magnetic field lines and the gray-scale contour shows the CR pressure.
The top three panels show the time evolution for the case of $\kappa_{\|}=10^{-4}$,
and the bottom three panels for the case of $\kappa_{\|}=10.0$.
The time evolution of the magnetic field lines looks like almost the same
even if the values of $\kappa_{\|}$ are different.
On the other hand, the CR pressure distribution are different with different $\kappa_{\|}$ value.
In the case of $\kappa_{\|}=10^{-4}$, the CR pressure becomes stronger slightly at the
valley of the magnetic field lines as the time proceeds.
However, in the case of $\kappa_{\|}=10.0$, it shows no variation as the time proceeds.

To compare the results obtained from linear analysis and MHD simulations,
we examine the temporal variation of $V_x$ at a particular point.
Figure \ref{fig10} shows the time evolution of the absolute value $|V_x|$ at $(x,z)=(0.0,0.5)$.
The solid-line corresponds to the case of $\kappa_{\|}=10^{-4}$,
the dash-line corresponds to the case of $\kappa_{\|}=10.0$, the dotted-line
correspond to the power-law relation given by the linear analysis.
The solid-line and the dash-line almost completely overlap with each other and the two lines appear to be
one line in this scale.
The slope of these lines agrees well with the dotted-line from linear analysis.

\subsection{Numerical results of the differentially rotating cylinder model}
In the differentially rotating cylinder model, we calculate within the region extracted from $r$-$z$ plane
as shown in Figure~\ref{fig3}.
The size of this region is $1.0H_0\times 0.5H_0$, with $r\in[0.5H_0,1.5H_0]$ and $z\in[0.0H_0,0.5H_0]$.
The numerical grid resolution and the grid size are
$N_r=81$, $N_z=42$, and $\Delta r=0.125H_0$, $\Delta z=0.125H_0$.
We assume a symmetric boundary condition at $r=r_{\rm min}$, a free boundary condition at $r=r_{\rm max}$,
and a periodic boundary condition at $z=z_{\rm min}$, $z=z_{\rm max}$.
The initial equilibrium state is described in section~\ref{sec:equil-cyl1}.
To start the simulation, a small velocity perturbation is added to the region where the rotation velocity is not zero,
\begin{equation}\label{Vphi-sim-cyl}
  \delta V_\phi = -10^{-3}\times\cos(k_z z)\,.
\end{equation}
We choose $k_z=25.0$ as a reference to the result of linear analysis (see left panel of Figure~\ref{fig7}).
With this choice the analysis of the results of the MHD simulation is easier,
because we need to control just two waves inside the simulation box.

Similar to the shearing box model, we also study the two values of the CR diffusion coefficient,
$\kappa_{\|}=10^{-4}$, and $\kappa_{\|}=10.0$ as the representative values
in accordance with the result of linear analysis (see the right panel of Figure~\ref{fig7}).
Figure \ref{fig11} shows the time evolution of the distributions of the magnetic field and the CR pressure.
In the figure the white curves are the magnetic field lines and the gray-scale contour shows the CR pressure.
The top three panels show the time evolution for the case of $\kappa_{\|}=10^{-4}$,
and the bottom three panels for the case of $\kappa_{\|}=10.0$.


In the case of small diffusion coefficient $\kappa_{\|}=10^{-4}$, the growth of the instability is slow.
It is still rather insignificant around $t\sim 2.0$, and the instability starts to grow around $t=3.0$
(see upper panels of Figure!\ref{fig10}).
On the other hand, in the case of larger diffusion coefficient, the instability is already approaching its saturation around
$t\sim 3.0$ (lower panels of Figure~\ref{fig11}).
As the growth of the instability proceeds, the low CR-pressure region penetrates into the high CR-pressure region around $t\sim 3.0$.

In order to understand the mechanism causing different growth rate of MRI, we compared the case of
$\kappa_{\|}=10.0$ with the case of $\kappa_{\|}=0.01$.
They show similar growth process of the instability in magnetic fields except that the growth rates are different.
The left panels of Figure~\ref{fig12} shows the density (gray scale contour), velocity distribution (white arrows),
and a reference magnetic field line (white curve) for $\kappa_{\|}=0.01$ at $t=3.45$ and $\kappa_{\|}=10.0$ at $t=3.0$.
The black arrow at the top-right corner is half the unit velocity, the Keplerian rotation speed at $r=1.0$.
High density region is created where the MRI is growing strongly.
The right panels of Figure~\ref{fig12} shows the CR pressure distribution, the density distribution, and
the toroidal velocity distribution along a reference magnetic field line for $\kappa_{\|}=0.01$ and 10.0.
The CR pressure distribution differs significantly for different $\kappa_{\|}$.
For large $\kappa_{\|}$ the CR pressure becomes uniform along the magnetic field line,
while for small $\kappa_{\|}$ the CR pressure varies in sync with the plasma density.
Density attains its maximum at the region where the MRI is growing strongly,
and its value is higher for the larger $\kappa_{\|}$.
The toroidal velocity varies anti-sync with density, but the distributions for different
$\kappa_{\|}$ are more or less the same.

\section{SUMMARY AND DISCUSSION}\label{sec:summary}

We studied the MRI with the effect of CRs by linear stability analysis and MHD simulation.
We examined two different models: the shearing box model and the differentially rotating cylinder model.

\begin{figure}
\epsscale{1.0}
%
%
\plottwo{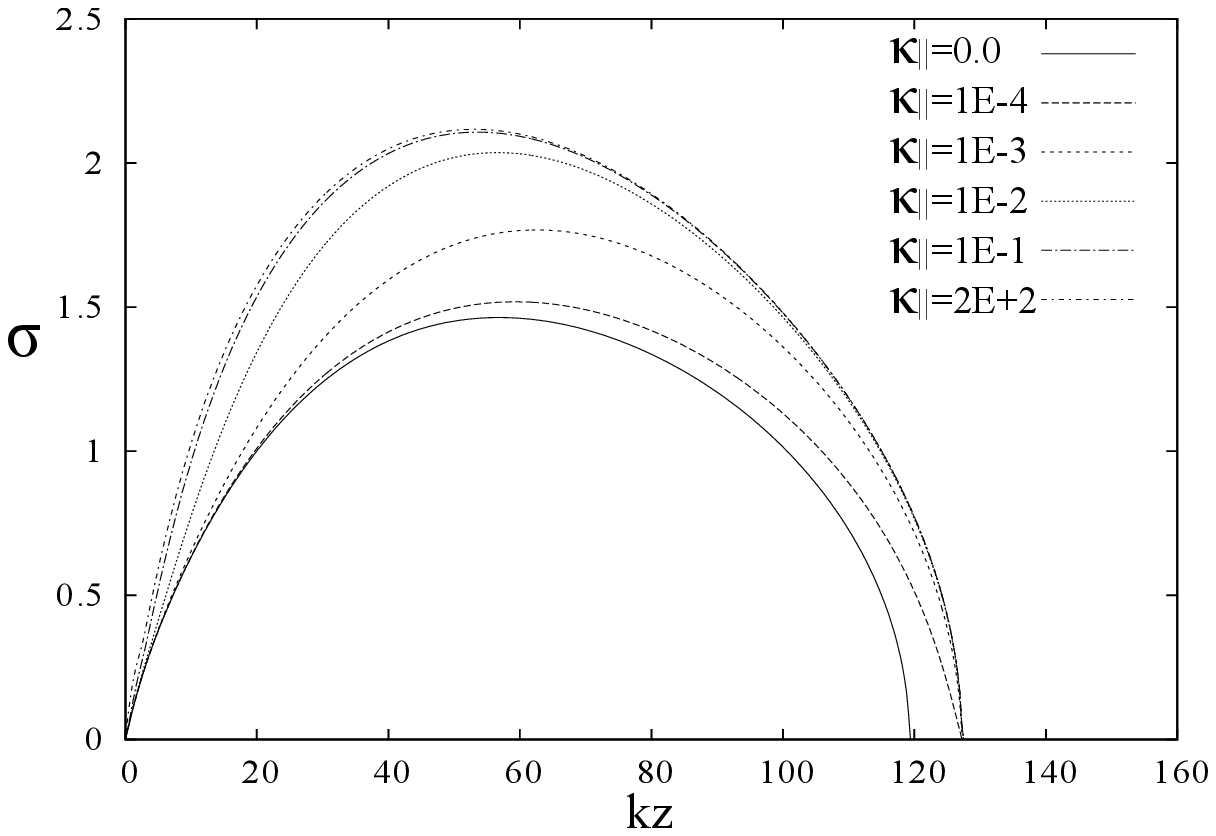}{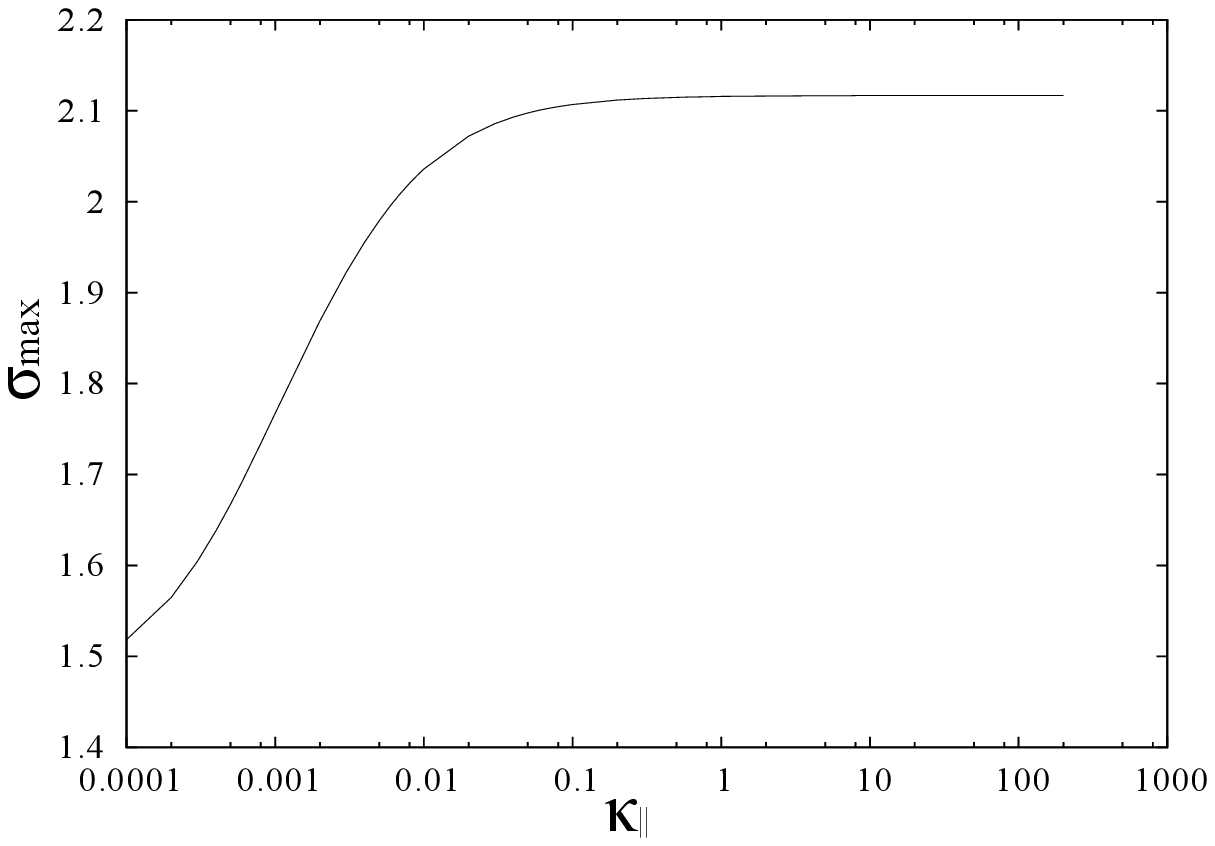}
\caption{
$Left$: Dispersion relation for the magnetorotational instability with the effect of CRs
at different $\kappa_{\|}$ in the differentially rotating cylinder model.
Here $\sigma$ is the growth rate of perturbation and $k_z$
is the wavenumber along the direction of the magnetic field in the unperturbed state.
$Right$: Dependence of the maximum growth rate $\sigma_{\rm max}$ on $\kappa_{\|}$.
$\sigma_{\rm max}$ is small when $\kappa_{\|}<0.0005$, it increases considerably
in the range $0.0005\le\kappa_{\|}\le 0.05$, and goes to saturation when $\kappa_{\|}>0.05$.
}
\label{fig7}
\end{figure}

\begin{figure}
\epsscale{1.0}
%
%
\plotone{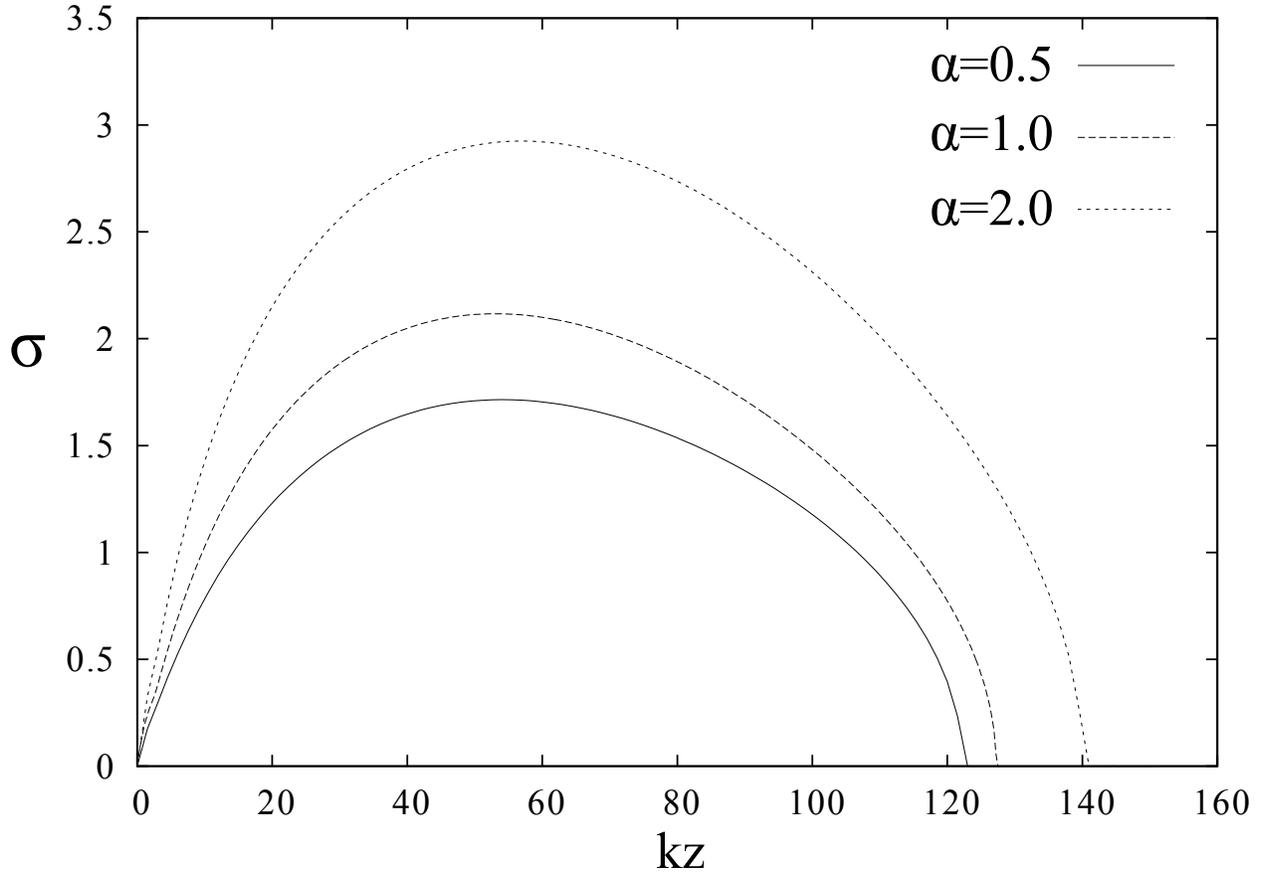}
\caption{
Same as the right panel of Figure~\ref{fig7} except that the CR diffusion coefficient is fixed at $\kappa_{\|}=200$ and
the ratio of CR pressure to thermal pressure $\alpha$ varies (while the sum of them is kept fixed).
}
\label{fig8}
\end{figure}

In linear stability analysis, we reduced the set of perturbation equation to two first order ODEs and
obtain the dispersion relation using shooting method.
For the shearing box model, the growth rate barely depends on the value of $\kappa_{\|}$ (see Figure~\ref{fig5}).
This is starkly different from previous studies on related topics \citep[e.g.,][]{ryu, kuwabara2004, kuwabara2006},
which showed considerably dependence of the growth rate on the value of $\kappa_{\|}$.
The reason lies in the distribution of CR pressure distribution in the initial unperturbed background.
If the CR pressure is uniform distributed in the unperturbed background (as in the case of the shearing box model),
then the growth rate will be (almost) independent of the value of $\kappa_{\|}$.
However, for non-uniform CR pressure distribution, the growth rate will depends on $\kappa_{\|}$.
We confirmed this in our second model, the differentially rotating cylinder model, which has a non-uniform
CR pressure distribution in the unperturbed background.
Figure~\ref{fig7} shows the dependence of the growth rate on $\kappa_{\|}$.
The growth rate increases as $\kappa_{\|}$ increases,
and saturated at large $\kappa_{\|}$ (see right panel of Figure~\ref{fig7} for the maximum growth rate).
This is consistent with the studies on Parker instability and Parker-Jeans instability
\citep{kuwabara2004, kuwabara2006}.
However, there are some subtle differences.
At small values of $\kappa_{\|}$ ($<0.001$), the maximum growth rate is more or less the same in MRI
(see right panel of Figure~\ref{fig7}), but this characteristics was not observed in the study of Parker instability
\citep{kuwabara2004}.
Figure~\ref{fig8} shows the dependence of the growth rate on the ratio of CR pressure to thermal pressure $\alpha$.
The growth rate increases as $\alpha$ increases samely in $\kappa_{\|}$.
An increase of $\alpha$ is equivalent to a decrease of the ratio of thermal pressure to magnetic pressure.
This result is somewhat different from the result by \citep[][]{khajenabi2012} that
the growth rate becomes larger as the ratio of thermal pressure to magnetic pressure is larger.
This difference is perhaps come from our formalism and the non-uniformity of the unperturbed state.
In our treatment (see Equation~\ref{iccyl3}) the density distribution is independent of $\alpha$ once we keep the
the sum of CR pressure and thermal pressure fixed. It is more convenient to study the effect of $\alpha$ without
changing the density profile.

In the MHD simulation for the shearing box model,
we also obtained the result that the growth rate of MRI does not depend on the $\kappa_{\|}$ (see Figure~\ref{fig9}).
In Figure~\ref{fig10}, we compared the growth rate obtained from the linear analysis with that obtained
from the MHD simulation, and they agreed well.
From these results (linear analysis and MHD simulations), we can conclude that
the growth of the MRI does not depend on the value of the CR diffusion coefficient $\kappa_{\|}$
when the initial background CR pressure distribution is uniform, at least in the linearly growing phase.

In the MHD simulation for the differentially rotating cylinder model, we find that
the growth rate of MRI under the non-uniform CR pressure background does depend on the value
of the CR diffusion coefficient $\kappa_{\|}$.
The growth of MRI becomes faster as the $\kappa_{\|}$ becomes larger (see Figure~\ref{fig11}).
This result is consistent with that obtained from the linear stability analysis.
This result shows that the MRI with cosmic-ray diffusion strongly depends on the distribution of the CR pressure
background.
If the distribution of CRs is non-uniform, the growth rate of MRI may change drastically with the value of $\kappa_{\|}$.


In the differentially rotating cylinder model,
the dependence of the MRI growth rate on the value of $\kappa_{\|}$ is caused by the difference in
CR pressure distribution along a magnetic field line.
A general property of diffusion is to smooth out irregularities and to reduce the gradient of the relevant quantity.
If the diffusion coefficient is large (i.e., weak coupling between plasma and CR), the CR pressure (or CR energy)
approaches uniform distribution quickly even if it were driven away from uniformity by the growth of MRI.
Under such circumstances, the CR pressure gradient along a magnetic field line becomes small and is not able to
curb the outward movement of plasma by the centrifugal force.
Consequently, high density region is formed at the location where MRI is growing and the magnetic field line
develops the loop like structure.
If the diffusion coefficient is small, the CR pressure maintains non-uniformity longer and hinders the outward
movement of the plasma.
Hence the density is smaller at the location where MRI is growing when compare with the large diffusion coefficient case.
On the other hand, the toroidal velocity distribution is not sensitive to the value of $\kappa_{\|}$
(see right panels of Figure~\ref{fig12}).
This means that the depicted magnetic field line in the case of small or large diffusion coefficient
($\kappa_{\|}$=0.01 or 10.0) rotates with the same rotation speed profile.
Therefore, the centrifugal force becomes larger at the higher density region and the growth rate becomes larger.

From these results, we speculate that the effect of CRs on MRI will be weak in the phase that
the turbulence is sufficiently grown up and the distribution of CR pressure approaches uniform.
Only in the phase when the turbulence is still growing and the CR pressure is non-uniform will the effect
of CRs on MRI become significant.



\acknowledgments

CMK is supported, in part, by the Taiwan Ministry of Science and Technology
grant MOST 102-2112-M-008-019-MY3.

\appendix

\section{Perturbation quantities}\label{sec:perturb-quantities}
As mentioned in the main text, the set of perturbation equations can be reduced to two first order ODEs
of $\delta\bar{V_x}$ and $\delta\bar{P_{\rm t}}$ in the case of shearing box model, and
$\delta\bar{V_r}$ and $\delta\bar{P_{\rm t}}^\prime$ in the case of differentially rotating cylinder model.
The other quantities are related to these two quantities algebraically.
We list them here explicitly.

\subsection{Shearing box}
First, we express $\delta\bar{P_{\rm g}}$, $\delta\bar{P_{\rm c}}$, $\delta\bar{B_x}$, $\delta\bar{B_y}$, $\delta\bar{B_z}$,
$\delta\bar{V_y}$ and $\delta\bar{V_z}$ in terms of $\delta\bar{V_x}$, $\delta\bar{P_{\rm t}}$ and $\delta{\bar\rho}$,
then $\delta{\bar\rho}$ in terms of $\delta\bar{V_x}$ and $\delta\bar{P_{\rm t}}$.
\begin{eqnarray}\label{dPg-shear}
  \delta\bar{P_{\rm g}}={1\over\Sigma}\left(C_{\rm s}^2{d\rho\over dx}-{d P_{\rm g}\over dx}\right)\delta\bar{V_x}
  +\rho C_{\rm s}^2{\delta{\bar\rho}\over\rho}\,,
\end{eqnarray}
\begin{eqnarray}\label{dPc-shear}
  \delta\bar{P_{\rm c}}={1\over\Sigma}\left[{C_{\rm c}^2\over(1+K)}{d\rho\over dx}
  -{d P_{\rm c}\over dx}\right]\delta\bar{V_x}
  +{\rho C_{\rm c}^2\over(1+K)}{\delta{\bar\rho}\over\rho}\,,
\end{eqnarray}
\begin{eqnarray}\label{dBx-shear}
  \delta\bar{B_x}=-\,{1\over\Sigma}\left(k_y B_y+k_z B_z\right)\delta\bar{V_x}\,,
\end{eqnarray}
\begin{eqnarray}\label{dBy-shear}
  \delta\bar{B_y}&=&{1\over(1+W)\Sigma}\left\{
  {1\over\Sigma}k_y\left(k_y B_y+k_z B_z\right){\delta\bar{P_{\rm t}}\over\rho}
  +\Sigma B_y{\delta{\bar\rho}\over\rho} \right. \nonumber \\
  &&\quad \left. +\left\{B_y\left[{1\over\rho}{d\rho\over dx}-{(1+W)\over B_y}{d B_y\over dx}\right]
  -i\,{2\Omega\over\Sigma}\left(k_y B_y+k_z B_z\right)\right\}
  \delta\bar{V_x}\right\}\,,
\end{eqnarray}
\begin{eqnarray}\label{dBz-shear}
  \delta\bar{B_z}&=&{1\over(1+W)\Sigma}\left\{
  {1\over\Sigma}k_z\left(k_y B_y+k_z B_z\right){\delta\bar{P_{\rm t}}\over\rho}
  +\Sigma B_z{\delta{\bar\rho}\over\rho} \right. \nonumber \\
  &&\quad \left. +B_z\left[{1\over\rho}{d\rho\over dx}-{(1+W)\over B_z}{d B_z\over dx}\right]
  \delta\bar{V_x}\right\}\,,
\end{eqnarray}
\begin{eqnarray}\label{dVy-shear}
  \delta\bar{V_y}&=&{1\over(1+W)\Sigma}\left\{
  -k_y{\delta\bar{P_{\rm t}}\over\rho}
  +{V_{\rm A}^2 B_y\left(k_y B_y+k_z B_z\right)\over\left(B_y^2+B_z^2\right)}{\delta{\bar\rho}\over\rho} \right. \nonumber \\
  &&\quad \left. +\left[{V_{\rm A}^2 B_y\left(k_y B_y+k_z B_z\right)\over\Sigma\left(B_y^2+B_z^2\right)\rho}{d\rho\over dx}
  +i\,(1+W){d V_y\over dx}+i\, 2\Omega\right]
  \delta\bar{V_x}\right\}\,,
\end{eqnarray}
\begin{eqnarray}\label{dVz-shear}
  \delta\bar{V_z}&=&{1\over(1+W)\Sigma}\left\{
  -k_z{\delta\bar{P_{\rm t}}\over\rho}
  +{V_{\rm A}^2 B_z\left(k_y B_y+k_z B_z\right)\over\left(B_y^2+B_z^2\right)}{\delta{\bar\rho}\over\rho} \right. \nonumber \\
  &&\quad \left. +\left[{V_{\rm A}^2 B_z\left(k_y B_y+k_z B_z\right)\over\Sigma\left(B_y^2+B_z^2\right)\rho}{d\rho\over dx}
  +i\,(1+W){d V_z\over dx}\right]
  \delta\bar{V_x}\right\}\,,
\end{eqnarray}
and finally,
\begin{eqnarray}\label{drho-shear}
  {\delta{\bar\rho}\over\rho}&=&{1\over{\cal A}^2}\left\{
  {1\over(1+W)}{\delta\bar{P_{\rm t}}\over\rho}
  +{1\over\Sigma}\left[{1\over\rho}{d P_{\rm t}\over dx}
  -{{\cal A}^2\over\rho}{d\rho\over dx} \right. \right. \nonumber \\
  &&\quad \left. \left. +i\,{2\Omega V_{\rm A}^2 B_y\left(k_y B_y+k_z B_z\right)\over(1+W)\Sigma\left(B_y^2+B_z^2\right)}
  \right]
  \delta\bar{V_x}\right\}\,.
\end{eqnarray}
The other quantities are given by Equations~(\ref{pert-eq-shear})-(\ref{Pt-shear}).

\subsection{Differentially rotating cylinder}
Similarly, we express $\delta\bar{P_{\rm g}}$, $\delta\bar{P_{\rm c}}$, $\delta\bar{B_r}$, $\delta\bar{B_\phi}$, $\delta\bar{B_z}$,
$\delta\bar{V_\phi}$ and $\delta\bar{V_z}$ in terms of $\delta\bar{V_r}$, $\delta\bar{P_{\rm t}}^\prime$ and $\delta{\bar\rho}$,
then $\delta{\bar\rho}$ in terms of $\delta\bar{V_x}$ and $\delta\bar{P_{\rm t}}^\prime$.
\begin{eqnarray}\label{dPg-cyl}
  \delta\bar{P_{\rm g}}={1\over\Sigma^\prime}\left(C_{\rm s}^2{d\rho\over dr}-{d P_{\rm g}\over dr}\right)\delta\bar{V_r}
  +\rho C_{\rm s}^2{\delta{\bar\rho}\over\rho}\,,
\end{eqnarray}
\begin{eqnarray}\label{dPc-cyl}
  \delta\bar{P_{\rm c}}={1\over\Sigma^\prime}\left[{C_{\rm c}^2\over(1+K^\prime)}{d\rho\over dr}
  -{d P_{\rm c}\over dr}\right]\delta\bar{V_r}
  +{\rho C_{\rm c}^2\over(1+K^\prime)}{\delta{\bar\rho}\over\rho}\,,
\end{eqnarray}
\begin{eqnarray}\label{dBx-cyl}
  \delta\bar{B_r}=-\,{1\over\Sigma^\prime}\left({m\over r} B_\phi+k_z B_z\right)\delta\bar{V_r}\,,
\end{eqnarray}
\begin{eqnarray}\label{dBphi-cyl}
  \delta\bar{B_\phi}&=&{1\over(1+W^\prime)\Sigma^\prime}\left\{
  {m\over\Sigma^\prime r}\left({m\over r} B_\phi+k_z B_z\right){\delta\bar{P_{\rm t}}^\prime\over\rho}
  +\Sigma^\prime B_\phi{\delta{\bar\rho}\over\rho} \right. \nonumber \\
  &&\quad \left. +\left\{B_\phi\left[{1\over\rho}{d\rho\over dr}-{(1+W^\prime)\over B_\phi}{d B_\phi\over dr}
  +{(1+W^\prime)\over r}\right] \right. \right. \nonumber \\
  &&\quad \left. \left. -i\,{2\Omega\over\Sigma^\prime}\left({m\over r} B_\phi+k_z B_z\right)\right\}
  \delta\bar{V_r}\right\}\,,
\end{eqnarray}
\begin{eqnarray}\label{dBz-cyl}
  \delta\bar{B_z}&=&{1\over(1+W^\prime)\Sigma^\prime}\left\{
  {k_z\over\Sigma^\prime}\left({m\over r} B_\phi+k_z B_z\right){\delta\bar{P_{\rm t}}^\prime\over\rho}
  +\Sigma^\prime B_z{\delta{\bar\rho}\over\rho} \right. \nonumber \\
  &&\quad \left. +B_z\left[{1\over\rho}{d\rho\over dr}-{(1+W^\prime)\over B_z}{d B_z\over dr}\right]
  \delta\bar{V_r}\right\}\,,
\end{eqnarray}
\begin{eqnarray}\label{dVphi-cyl}
  \delta\bar{V_\phi}&=&{1\over(1+W^\prime)\Sigma^\prime}\left\{
  -\,{m\over r}{\delta\bar{P_{\rm t}}^\prime\over\rho}
  +{{V^\prime_{\rm A}}^2 B_\phi\over\left(B_\phi^2+B_z^2\right)}\left({m\over r} B_\phi+k_z B_z\right)
  {\delta{\bar\rho}\over\rho} \right. \nonumber \\
  &&\quad \left. +\left[{{V^\prime_{\rm A}}^2 B_\phi
  \over\Sigma^\prime\left(B_\phi^2+B_z^2\right)}\left({m\over r} B_\phi+k_z B_z\right)
  \left({1\over\rho}{d\rho\over dr}+{2\over r}\right) \right. \right. \nonumber \\
  &&\quad \left. \left. +i\,(1+W^\prime)r{d\Omega\over dr}+i\,2\Omega\right]
  \delta\bar{V_r}\right\}\,,
\end{eqnarray}
\begin{eqnarray}\label{dVz-cyl}
  \delta\bar{V_z}&=&{1\over(1+W^\prime)\Sigma^\prime}\left\{
  -k_z{\delta\bar{P_{\rm t}}^\prime\over\rho}
  +{{V^\prime_{\rm A}}^2 B_z\over\left(B_\phi^2+B_z^2\right)}\left({m\over r} B_\phi+k_z B_z\right)
  {\delta{\bar\rho}\over\rho} \right. \nonumber \\
  &&\quad \left. +\left[{{V^\prime_{\rm A}}^2 B_z
  \over\Sigma^\prime\left(B_\phi^2+B_z^2\right)\rho}{d\rho\over dr}\left({m\over r} B_\phi+k_z B_z\right)
  +i\,(1+W^\prime){d V_z\over dr}\right]
  \delta\bar{V_r}\right\}\,,
\end{eqnarray}
and finally,
\begin{eqnarray}\label{drho-cyl}
  {\delta{\bar\rho}\over\rho}&=&{1\over{\cal A^\prime}^2}\left\{
  {1\over(1+W^\prime)}{\delta\bar{P_{\rm t}}^\prime\over\rho}
  +{1\over\Sigma^\prime}\left[{1\over\rho}{d P_{\rm t}^\prime\over dr}
  -{{\cal A^\prime}^2\over\rho}{d\rho\over dr}
  -\,{(1-W^\prime){V^\prime_{\rm A}}^2 B_\phi^2\over(1+W^\prime) r\left(B_\phi^2+B_z^2\right)} \right. \right. \nonumber \\
  &&\quad \left. \left. +i\,{2\Omega {V^\prime_{\rm A}}^2 B_\phi
  \over(1+W^\prime)\Sigma^\prime\left(B_\phi^2+B_z^2\right)}\left({m\over r} B_\phi+k_z B_z\right)
  \right]
  \delta\bar{V_r}\right\}\,.
\end{eqnarray}
The other quantities are given by Equations~(\ref{pert-eq-cyl})-(\ref{Pt-cyl}).

\clearpage



\clearpage

\clearpage

\clearpage

\clearpage

\clearpage

\clearpage

\clearpage

\clearpage

\begin{figure}
\epsscale{0.8}
%
%
\plotone{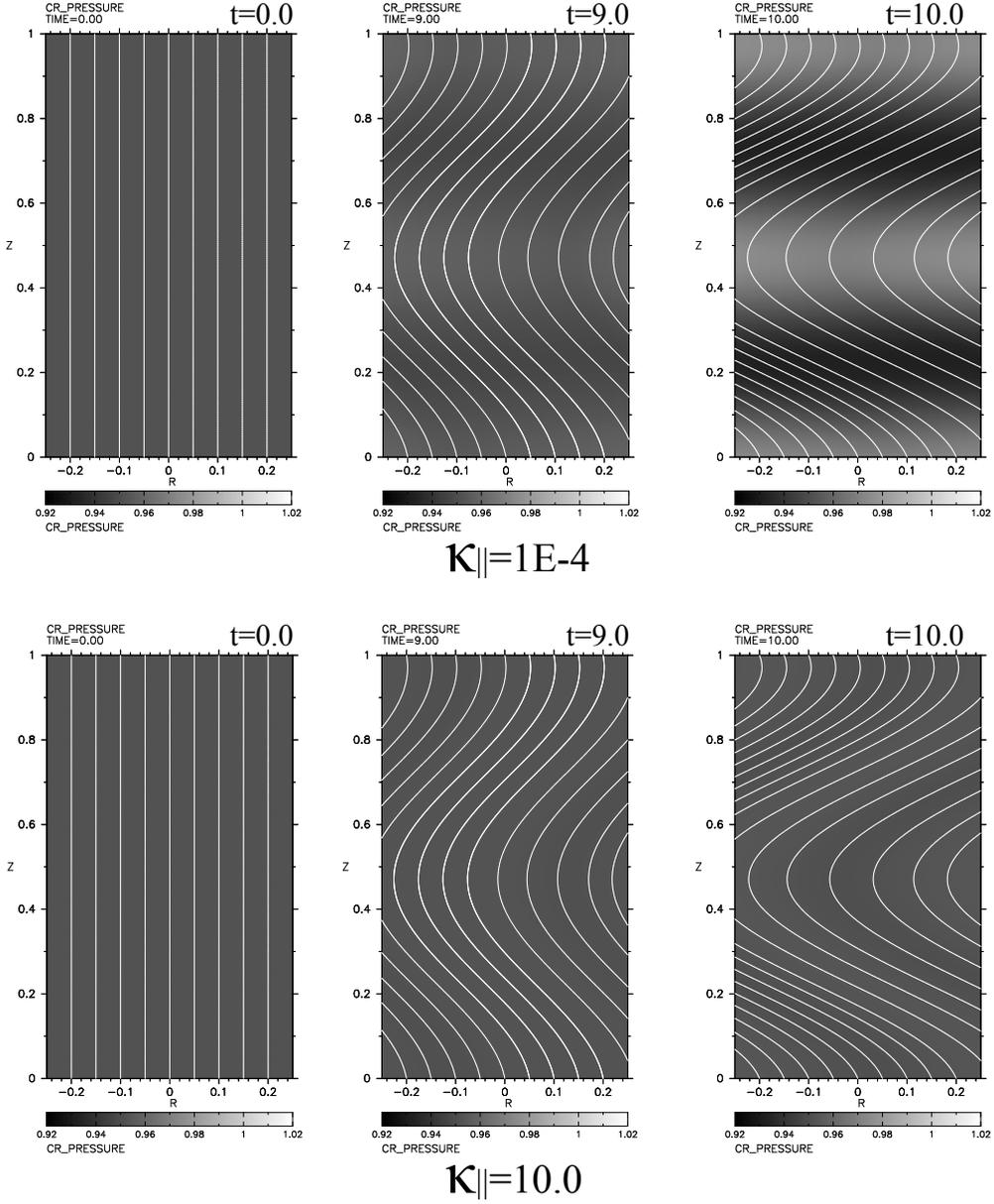}
\caption{
Time evolution of the CR pressure distribution and magnetic field lines of the shearing box model
for the cases of $\kappa_{\|}=10^{-4}$ ($top$) and $10.0$ ($bottom$).
The gray scale and the white curves show the CR pressure distribution and magnetic field lines, respectively.
The magnetic field lines behave almost the same in both cases.
However, as time proceeds, the CR pressure becomes slightly larger at the valley of the magnetic field lines
for the small $\kappa_{\|}$ case.
}
\label{fig9}
\end{figure}

\clearpage

\begin{figure}
\epsscale{1.0}
%
%
\plotone{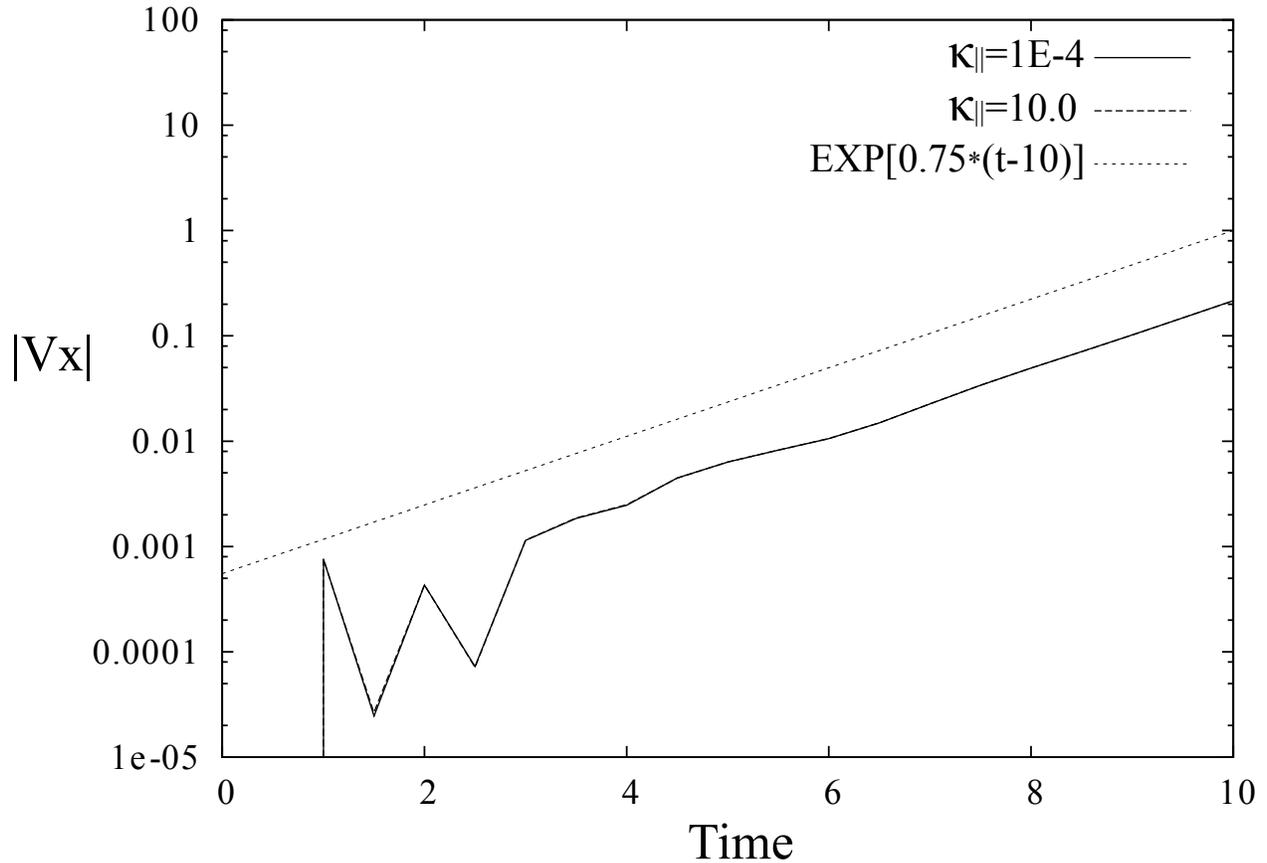}
\caption{
Comparison of the growth rate obtained from MHD simulations and linear analysis in the shearing box model.
The results of the two simulations almost overlap each other completely (it is difficult to distinguish them in this scale).
This verifies the conclusion of the result of the linear analysis (see Figure~\ref{fig5}).
The line $\exp[0.75*(t-10.0)]$ shows the power-law relation given by the linear analysis.
The result of MHD simulations and linear analysis agree well with each other.
}
\label{fig10}
\end{figure}

\clearpage

\begin{figure}
\epsscale{1.0}
%
%
\plotone{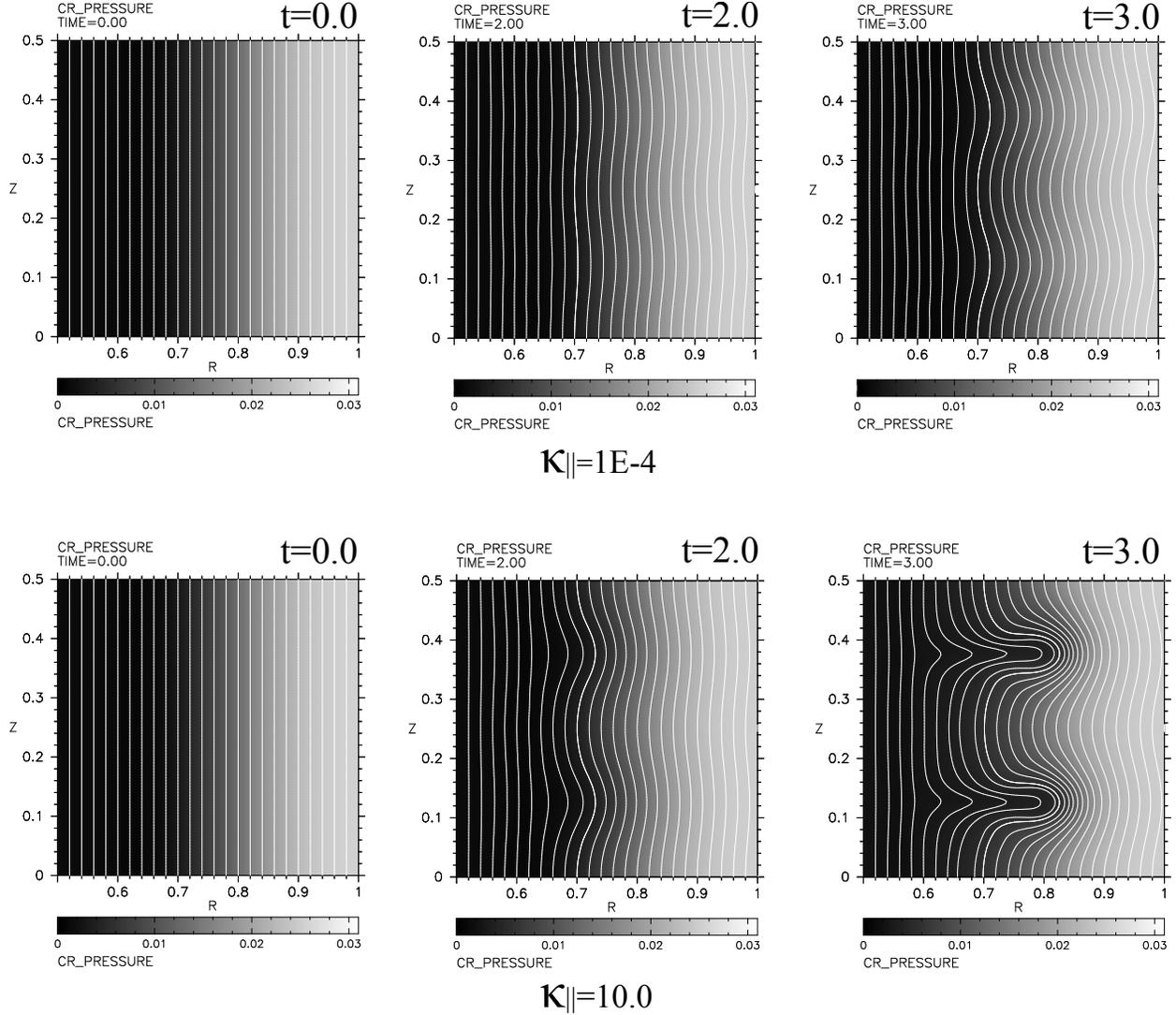}
\caption{
Time evolution of the CR pressure distribution and magnetic field lines of the differentially
rotating cylinder model for the cases of $\kappa_{\|}=10^{-4}$ ($top$) and $10.0$ ($bottom$).
The gray scale and the white curves show the CR pressure distribution and magnetic field lines, respectively.
The growth of the instability is slower in the case of small diffusion coefficient when compare to the large
diffusion coefficient one.
}
\label{fig11}
\end{figure}

\clearpage

\begin{figure}
\epsscale{1.0}
%
%
\plotone{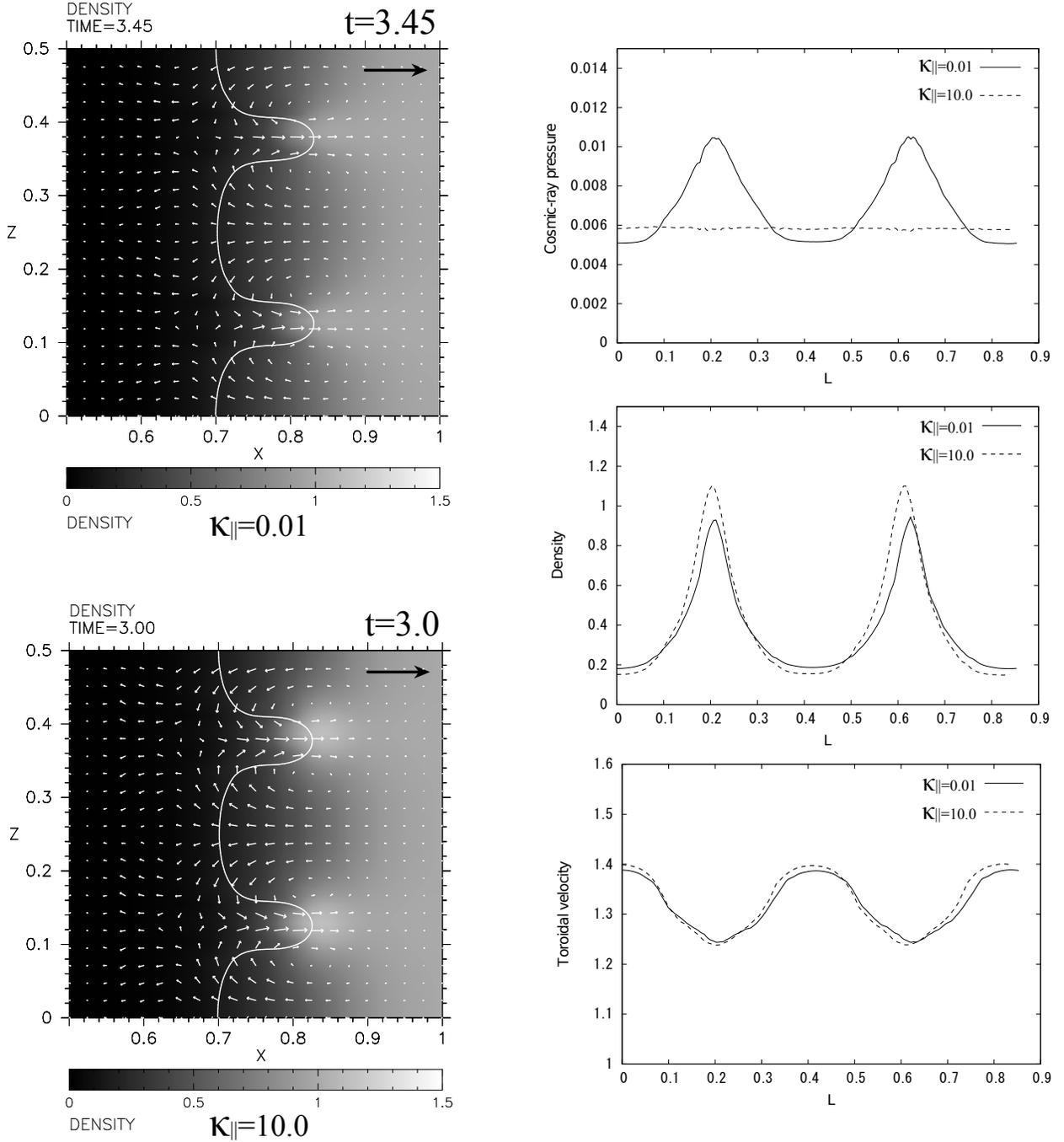}
\caption{
$Left$: The distributions of CR pressure (gray scale), toroidal velocity (white arrows) and a reference
magnetic field line (white curve) for the case $\kappa_{\|}=0.01$ at $t=3.45$ and the case $\kappa_{\|}=10.0$ at $t=3.0$.
The black arrow is half the Keplerian rotation speed at $r=1.0$.
$Right$: The CR pressure, density and toroidal velocity along the reference magnetic field line (depicted in the left panels).
$L$ is the distance along the magnetic field line.
Solid line is the case $\kappa_{\|}=0.01$ and dashed line is $\kappa_{\|}=10.0$.
}
\label{fig12}
\end{figure}

\end{document}